\documentclass[prc,superscriptaddress,showpacs,amsmath,amssymb,twocolumn]{revtex4-1}

\usepackage{amsmath, amsthm, amssymb, braket}

\usepackage{amsfonts}

\usepackage{graphicx}
\usepackage{dcolumn}
\usepackage{bm}

\usepackage{color}

\newcommand{\be}{\begin{equation}}
\newcommand{\ee}{\end{equation}}
\newcommand{\dd}{\mathrm{d}}

\newcommand{\ii}{\mathrm{i}}
\usepackage{ulem}

\begin{document}

\title{Charge-exchange dipole excitations in deformed nuclei
}

\author{Kenichi Yoshida}
\affiliation{Department of Physics, Kyoto University, Kyoto, 606-8502, Japan}

\date{\today}

\begin{abstract}
\noindent
\textbf{Background:} The electric giant-dipole resonance (GDR) is the most established collective vibrational mode of excitation. 
A charge-exchange analog, however, has been poorly studied in comparison with the spin (magnetic) dipole resonance (SDR).
\\
\textbf{Purpose:} I investigate the role of deformation 
on the charge-exchange dipole excitations 
and explore the generic features as an isovector mode of excitation. 
\\
\textbf{Methods:} The nuclear energy-density functional method is employed for calculating the response functions 
based 
on the Skyrme--Kohn--Sham--Bogoliubov method and the proton-neuton quasiparticle-random-phase approximation. 
\\
\textbf{Results:} The deformation splitting into $K=0$ and $K=\pm 1$ components occurs in the charge-changing channels 
and is proportional to the magnitude of deformation as is well known for the GDR. 
For the SDR, however, 
a simple assertion based on geometry of a nucleus cannot be applied for explaining the vibrational frequencies of each $K$-component. 
A qualitative argument on the strength distributions for each component is given based on the 
non-energy-weighted sum rules taking nuclear deformation into account. 
The concentration of the electric dipole strengths in low energy and below the giant resonance is found in neutron-rich unstable nuclei. 
\\
\textbf{Conclusions:}  The deformation splitting occurs generically for the charge-exchange dipole excitions as in the neutral channel. 
The analog pygmy dipole resonance can emerge in deformed neutron-rich nuclei as well as in spherical systems. 
\end{abstract}

\maketitle

\section{Introduction}\label{intro}

Charge-exchange excitations dovetail with the transitions 
from a mother nucleus $(Z,A)$ with proton number $Z$ and total nucleon number $A$ 
to final states in a neighboring daughter $(Z\pm 1,A)$ 
in the isospin lowering $\tau_-$ and raising $\tau_+$ channels, respectively.
They take place either in the charged-current nuclear (semileptonic) weak processes such as 
the $\beta$-decay, charged lepton capture and neutrino-nucleus reactions 
or in the hadronic reactions of $(p,n)$ or $(n,p)$ type. 
Therefore, the spin-isospin responses induced by the charge-exchange excitations 
present active and broad research topics in 
the fields of fundamental physics~\cite{ost92,eji00,mea01,lan03,ich06,avi08,fuj11,nak17,len19}.

Response of a nucleus unveils elementary modes of excitation emerged by the interactions and correlations 
among constituent nucleons.  
The nuclear response is characterized by the transferred angular momentum $\Delta L$, spin $\Delta S$ and isospin $\Delta T$~\cite{har01}.
The isovector (IV) giant dipole resonance (GDR) represented as $\Delta L=1, \Delta S=0, \Delta T=1$
is one of the well studied collective vibrational modes of excitation 
among various types of giant resonance~\cite{ber75}. 
The GDR is an out-of-phase spatial oscillation of protons and neutrons, and thus represented as $\Delta T_z=0$. 
Recently with the advent of RI-beam technology, 
a considerable amount of work has been devoted to a quest for exotic modes of excitation in nuclei far from the $\beta$-stability line 
and the low-energy dipole (LED) mode or the pygmy dipole resonance (PDR) 
has attracted a lot of interest~\cite{par07,aum13,roc18,bra19}. 
Furthrermore, the photoresonance can be seen in a wider perspective 
when it is considered as a single component $\Delta T_z=0$ of the IV dipole modes~\cite{BM2,izu83,aue84}. 
The additional components $\Delta T_z= \pm 1$ represent the charge-exchange modes.

Importance of the higher multipole spin-isospin responses beyond the allowed transitions is recognized 
for the weak processes in stellar environment~\cite{lan03,eji19}. 
The forbidden transitions are also involved
at zero temperature 
such as 
the neutrinoless double $\beta$-decay~\cite{avi08,eji19} if any and 
the single $\beta$-decay of heavy neutron-rich nuclei~\cite{bor03,suz12,zhi13,fan13,mus16,mar16,yos19}. 
From a nuclear structure point of view, 
the spin-dipole resonance (SDR) with $\Delta L=1,\Delta S=1, \Delta T=1$ 
have been studied to understand the mechanism for the collectivity of a giant resonance and 
the spin-isospin part of the interaction in nuclear medium~\cite{har01,ost92} 
besides that the Gamow--Teller and M1 resonances have been extensively studied as the IV magnetic $\Delta L=0$ transitions~\cite{fuj11}. 
The multipole dependence of the SDR elucidates 
the characteristic effects of the tensor force~\cite{bai10,hor13}, and 
the strengths of  
the dipole resonance are correlated with the neutron-skin thickness~\cite{roc18}. 
Though the study on the charge-exchange electric dipole resonance is limited, 
a recent work investigated 
a possible appearance of 
an analog of the PDR in neutron-rich nuclei, 
and the low-lying excitation corresponding to $-1\hbar \omega_0$ 
in very neutron-rich nuclei~\cite{yos17}.

In this article, I am going to investigate the deformation effects 
on the charge-exchange dipole resonances with both $\Delta S=0$ and $1$, 
and explore the generic features of dipole resonances as an IV mode of excitation. 
Furthermore, the roles of neutron excess is studied in details for the electric excitations and 
a possible appearance of the low-lying states is discussed. 
The present study is considered as an extension of the previous work on spherical nuclei~\cite{yos17} to deformed cases. 
From light to heavy nuclei are taken as a target of investigation to extract universal features associated with nuclear deformation. 
To this end, I employ 
the nuclear energy-density functional (EDF) method, which is a theoretical model being capable of 
handling nuclides with arbitrary mass number in a single framework~\cite{ben03,nak16}.

This paper is organized in the following way: 
The theoretical framework for describing the ground state of a mother nucleus, 
the excited states of a daughter nucleus and the transitions between them is given in Sec.~\ref{method} and 
details of the numerical calculation are also given; 
Sec.~\ref{result} is devoted to the numerical results and discussion based on the model calculation; 
the electric dipole resonance is studied in Sec.~\ref{result_dipole}, and the SDR in Sec.~\ref{result_SD}; 
then, summary is given in Sec.~\ref{summary}.

\section{Framework}\label{method}

\subsection{KSB and pnQRPA for deformed nuclei}

Since the details of the formalism can be found in Ref.~\cite{yos13}, 
here I show only the gist of the basic equations relevant to the present study.  
In a framework of the nuclear energy-density functional (EDF) method I employ, 
the ground state of a mother (target) nucleus is described by solving the 
Kohn--Sham--Bogoliubov (KSB) equation~\cite{dob84}:
\begin{align}
\begin{bmatrix}
h^{q}(\boldsymbol{r} \sigma)-\lambda^{q} & \tilde{h}^{q}(\boldsymbol{r} \sigma) \\
\tilde{h}^{q}(\boldsymbol{r} \sigma) & -h^{q}(\boldsymbol{r} \sigma)+\lambda^{q}
\end{bmatrix}
\begin{bmatrix}
\varphi^{q}_{1,\alpha}(\boldsymbol{r} \sigma) \\
\varphi^{q}_{2,\alpha}(\boldsymbol{r} \sigma)
\end{bmatrix} 
= E_{\alpha}
\begin{bmatrix}
\varphi^{q}_{1,\alpha}(\boldsymbol{r} \sigma) \\
\varphi^{q}_{2,\alpha}(\boldsymbol{r} \sigma)
\end{bmatrix}, \label{HFB_eq}
\end{align}
where 
the KS potentials $h$ and $\tilde{h}$ are given by the functional derivative of the EDF 
with respect to the particle density and the pair density, respectively. 
The superscript $q$ denotes 
$\nu$ (neutron, $ t_z= 1/2$) or $\pi$ (proton, $t_z =-1/2$). 
Assuming the system is axially symmetric, 
the KSB equation (\ref{HFB_eq}) is block diagonalized 
according to the quantum number $\Omega$, the $z$-component of the angular momentum. 

The excited states $| i \rangle$ of a daughter nucleus are described as 
one-phonon excitations built on the ground state $|0\rangle$ of the mother nucleus as 
\begin{align}
| i \rangle &= \hat{\Gamma}^\dagger_i |0 \rangle, \\
\hat{\Gamma}^\dagger_i &= \sum_{\alpha \beta}\left\{
X_{\alpha \beta}^i \hat{a}^\dagger_{\alpha,\nu}\hat{a}^\dagger_{\beta, \pi}
-Y_{\alpha \beta}^i \hat{a}_{\bar{\beta},\pi}\hat{a}_{\bar{\alpha},\nu}\right\},
\end{align}
where $\hat{a}^\dagger_\nu (\hat{a}^\dagger_\pi)$ and  $\hat{a}_\nu (\hat{a}_\pi)$ are 
the neutron (proton) quasiparticle (qp) creation and annihilation operators that 
are defined in terms of the solutions of the KSB equation (\ref{HFB_eq}) with the Bogoliubov transformation. 
The phonon states, the amplitudes $X^i, Y^i$ and the vibrational frequency $\omega_i$, 
are obtained in the proton-neutron quasiparticle-random-phase approximation (pnQRPA). 
The residual interactions entering into the pnQRPA equation 
are given by the EDF self-consistently. 
For the axially symmetric nuclei, the pnQRPA equation 
is block diagonalized according to the quantum number $K=\Omega_\alpha + \Omega_\beta$.

\subsection{Numerical procedures}

To describe the developed neutron skin and the neutrons pair correlation 
coupled with the continuum states that emerge uniquely in neutron-rich nuclei, 
I solve the KSB equation in the coordinate space using cylindrical coordinates
$\boldsymbol{r}=(\rho,z,\phi)$ with a mesh size of
$\Delta\rho=\Delta z=0.6$ fm and a box
boundary condition at $(\rho_{\mathrm{max}},z_{\mathrm{max}})=(14.7, 14.4)$ fm. 
Since I assume further the reflection symmetry, only the region of $z\geq 0$ is considered. 
The qp states are truncated according to the qp 
energy cutoff at 60 MeV, and 
the qp states up to the magnetic quantum number $\Omega=23/2$
with positive and negative parities are included. 
I introduce the truncation for the two-quasiparticle (2qp) configurations in the QRPA calculations,
in terms of the 2qp-energy as 60 MeV. 

For the normal (particle-hole) part of the EDF,
I employ mainly the SkM* functional~\cite{bar82}. 
For the pairing energy, I adopt the one in Ref.~\cite{yam09}
that depends on both
the isoscalar and isovector densities, 
in addition to the pair density, with the parameters given in
Table~III of Ref.~\cite{yam09}. 
The same pairing EDF is employed for the spin-singlet pn-pairing 
in the pnQRPA calculation, 
while the linear term in the isovector density is dropped. 
The same pairing strength is used for 
the dynamic spin-triplet pairing, 
though the occurrence of its condensation has been under active discussion~\cite{fra14}. 
Note that the pnQRPA calculations including the dynamic spin-triplet pairing with 
more or less the same strength as the spin-singlet pairing 
describe well the characteristic low-lying Gamow--Teller 
strength distributions in the light $N \simeq Z$ nuclei~\cite{fuj14,fuj15,fuj19}.

\section{Results and discussion}\label{result}

\subsection{Dipole excitations}\label{result_dipole}

\begin{figure}[t]
\begin{center}
\includegraphics[scale=0.3]{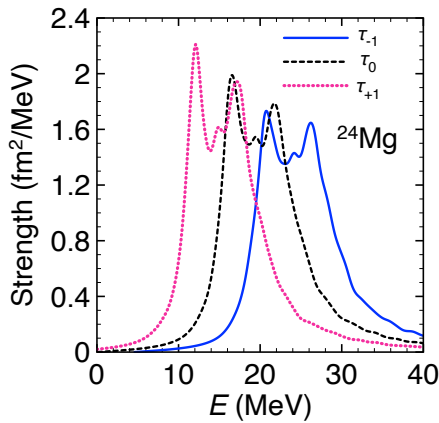}
\caption{
Calculated distributions of the IV dipole transition strengths by employing the SkM* functional 
as functions of the 
excitation energy with respect to the ground-state of $^{24}$Mg.
The smearing parameter $\gamma=2$ MeV is used. 
 }
\label{fig:24Mg}
\end{center}
\end{figure}

In Fig.~\ref{fig:24Mg}, presented are 
the transition-strength distributions in $^{24}$Mg as an example of deformed nuclei 
for the isovector dipole operators as functions of the 
excitation energy $E$ with respect to 
ground state of the mother (target) nucleus: 
\begin{align}
S^\pm (E)
&=\sum_{K}S^\pm_K (E) \\
&=\sum_{K}\sum_{i} \dfrac{\gamma/2}{\pi}\dfrac{ R^\pm_{i,K} }
{\{E- [ \hbar\omega_{i} \pm ( \lambda_\nu - \lambda_\pi)]\}^{2}+\gamma^{2}/4}, \label{response} \\
R^\pm_{i,K} &= | \langle i | \hat{F}^\pm_{1K} |0 \rangle  |^2 = |\langle 0 |[\hat{\Gamma}_i, \hat{F}^\pm_{1K}] |0 \rangle|^2,
\end{align} 
where $\lambda_\nu (\lambda_\pi)$ is the chemical potential for neutrons (protons) 
and the mass difference between a neutron and a proton is ignored.  
The charge-exchange dipole operators are defied as
\begin{equation}
\hat{F}^\pm_{1K} = \sum_{\sigma, \sigma^\prime, \tau, \tau^\prime}\int \dd\boldsymbol{r} 
rY_{1K}(\hat{r}) \langle \sigma |1|\sigma^\prime\rangle \langle \tau |\tau_{\pm 1}|\tau^\prime\rangle
\hat{\psi}^\dagger(\boldsymbol{r}  \sigma \tau)  \hat{\psi}(\boldsymbol{r}  \sigma^\prime \tau^\prime) \label{dipole_op}
\end{equation}
in terms of the nucleon field operators. 
Below, I call the $\Delta T_z=\pm 1$ excited state as that induced by the operator $\hat{F}^{\pm}$. 

A distinct feature seen in the deformed system is the appearance of two-humped peak structure. 
In spherical nuclei, the giant resonance has a single peak except 
the shoulder structure due to the pygmy resonance in neutron-rich nuclei 
as discussed in Ref.~\cite{yos17}. 
I am thus going to discuss the mechanism for the occurrence of the two-humped peak shape for the giant resonance. 
For reference, 
the strength distribution for the operator $\hat{F}^0_{1K}$ 
is also shown in Fig.~\ref{fig:24Mg}, 
calculated in the like-particle QRPA framework~\cite{yos08, yos13b}. 
Here, $\hat{F}^0_{1K}$ is defied for $\tau_0$ in Eq.~(\ref{dipole_op}), with 
$\tau_{\pm1, 0}$ being the spherical components of the nucleonic isospin: 
$\tau_{\pm1}=\mp\frac{1}{\sqrt{2}}(\tau_x \pm \ii \tau_y), \tau_0=\tau_z$.  
It is noted that the operator $\hat{F}^0_{1K}$ is different from 
the standard IV dipole operator~\cite{BM2,rin80}. 
We assumed here the $^{24}$Mg nucleus is unpaired due to the large deformed shell gap of 12. 
When the Coulomb interaction is discarded, 
the transition-strength distributions for $\tau_{\pm1}$ and $\tau_0$ are identical to one another 
because the ground-state isospin is zero. 
Therefore, the origin of the two-humped peak structure may be due to 
the $K$-splitting that can be seen in the photoabsorption cross sections of deformed nuclei~\cite{BM2}.  
However, the intuitive picture of the out-of-phase spatial oscillation of protons and neutrons cannot be applied  
to the charge-exchange dipole modes, and 
I am going to investigate further the roles of deformation in other systems below. 
When the Coulomb interaction is turned on, 
the chemical potential for protons becomes higher than that for neutrons; 
the difference is 4.61 MeV.  
The spatial distribution of neutrons are thus shrunk.   
These structure changes can be seen in the excitation energy and transition strengths.  
Let me briefly discuss this point before investigating the heavier systems. 

The unperturbed mean-excitation energy for the isovector modes with $\Delta T_z$ 
built on the $T_0=0$ state in an $N=Z$ nucleus can be given by 
\begin{equation}
E^{(0)} (\Delta T_z)\simeq E^{(0)}(\Delta T_z=0)-\Delta T_z \Delta E_{\rm{Coul}}, 
\end{equation}
where $\Delta E_{\rm{Coul}}$ is the shift of the Coulomb energy per unit $Z$~\cite{BM2}.  
In the present framework, 
the Coulomb-energy shift $\Delta E_{\rm{Coul}}$ is represented approximately 
as the difference of the chemical potentials 
of the mother nucleus $\lambda_\pi -\lambda_\nu$. 
The energy shifts of the $\Delta T_z=0$ and $\pm 1$ modes due to the RPA correlation 
are not very different from each other 
in a light $N=Z$ nucleus, and they are about 5.2 MeV in the present calculation. 
Thus, the excitation energy of the $\Delta T_z=+1$ mode is lower than that of the $\Delta T_z=-1$ mode 
due to the Coulomb-energy shift.

Let us then discuss the transition strengths. 
One sees from Fig.~\ref{fig:24Mg} that the transition strengths for the $\Delta T_z=+1$ excitation are 
larger than those for $\Delta T_z=-1$. 
Since the strengths are concentrated into the giant resonance, 
one can apply the argument based on the sum rule to 
the qualitative understanding of the imbalanced strengths. 
The model-independent sum rule for the charge-exchange dipole modes 
in an axially deformed nucleus is given as 
\begin{align}
&\int \dd E [S^-_K(E)-S^+_K(E)] \notag \\
&=
\left\{
\begin{aligned}
2\dfrac{3}{4\pi}[N\langle z^2 \rangle_N - Z\langle z^2\rangle_Z]& \,\,\, (K=0) \\
2\dfrac{3}{8\pi}[N\langle \rho^2 \rangle_N - Z\langle \rho^2\rangle_Z]& \,\,\, (K=\pm1) 
\end{aligned}
\right.,
\label{eq:sum_rule}
\end{align}
where $\langle \cdot \rangle_{N(Z)}$ denotes the expectation value for neutrons (protons) 
and a factor of two comes from isospin with the definition of Eq.~(\ref{dipole_op}).
In the spherical limit, the sum rule value for each $K$ component coincides 
with  $\frac{1}{2\pi}[N\langle r^2\rangle_N - Z\langle r^2\rangle_Z]$.  
 In the present case, $N=Z$, the difference in the transition strengths for the $\Delta T_z=\pm1$ modes 
comes from the difference in the spatial extension of protons and neutrons; 
the calculated
root-mean-square radius of protons and neutrons is 
3.03 fm and 2.99 fm.

Furthermore, one sees that an average of the transition strengths of the $\Delta T_z=\pm1$ modes is close to 
that of the $\Delta T_z=0$ mode. 
A simple RPA analysis for a single normal mode employing the separable dipole interaction 
gives the relation for the transition strengths as 
\begin{equation}
\dfrac{1}{2}(S^-+S^+)=\left[1+O\left(\frac{N-Z}{A}\right)\right]S^0,
\label{eq:strength_rel}
\end{equation}
where $S^0$ is the transition strength of the $\Delta T_z=0$ mode~\cite{BM2}.  
Though this relation is model dependent, 
the present self-consistent calculation based on the nuclear EDF 
satisfies it within 2 $\%$ accuracy for each $K$. 
This implies that the giant resonances calculated here are collective; 
the microscopically computed giant resonance in $^{24}$Mg can be viewed as a single mode.   

 \begin{figure*}[t]
\begin{center}
\includegraphics[scale=0.32]{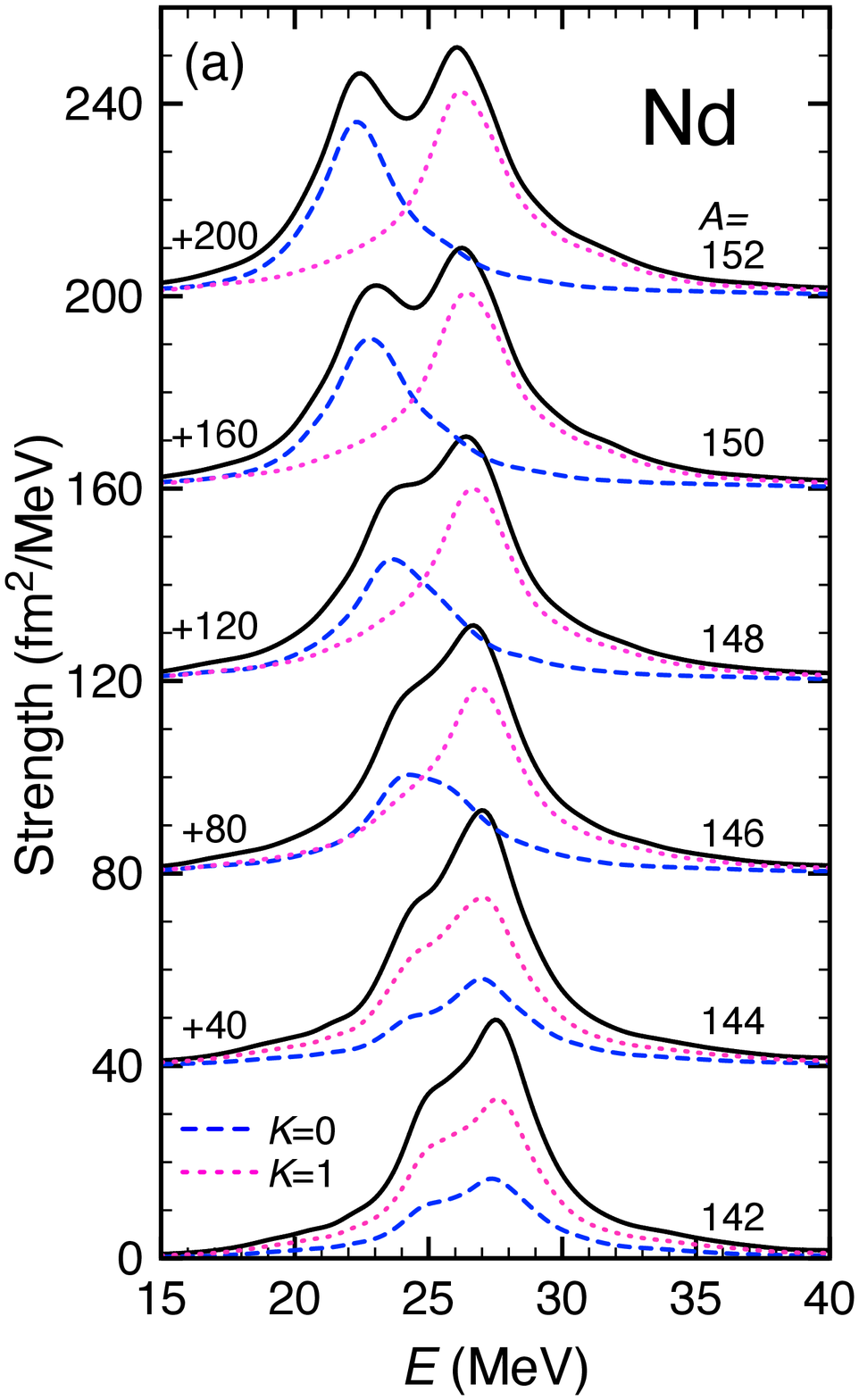}\!\!\!\!\!
\includegraphics[scale=0.32]{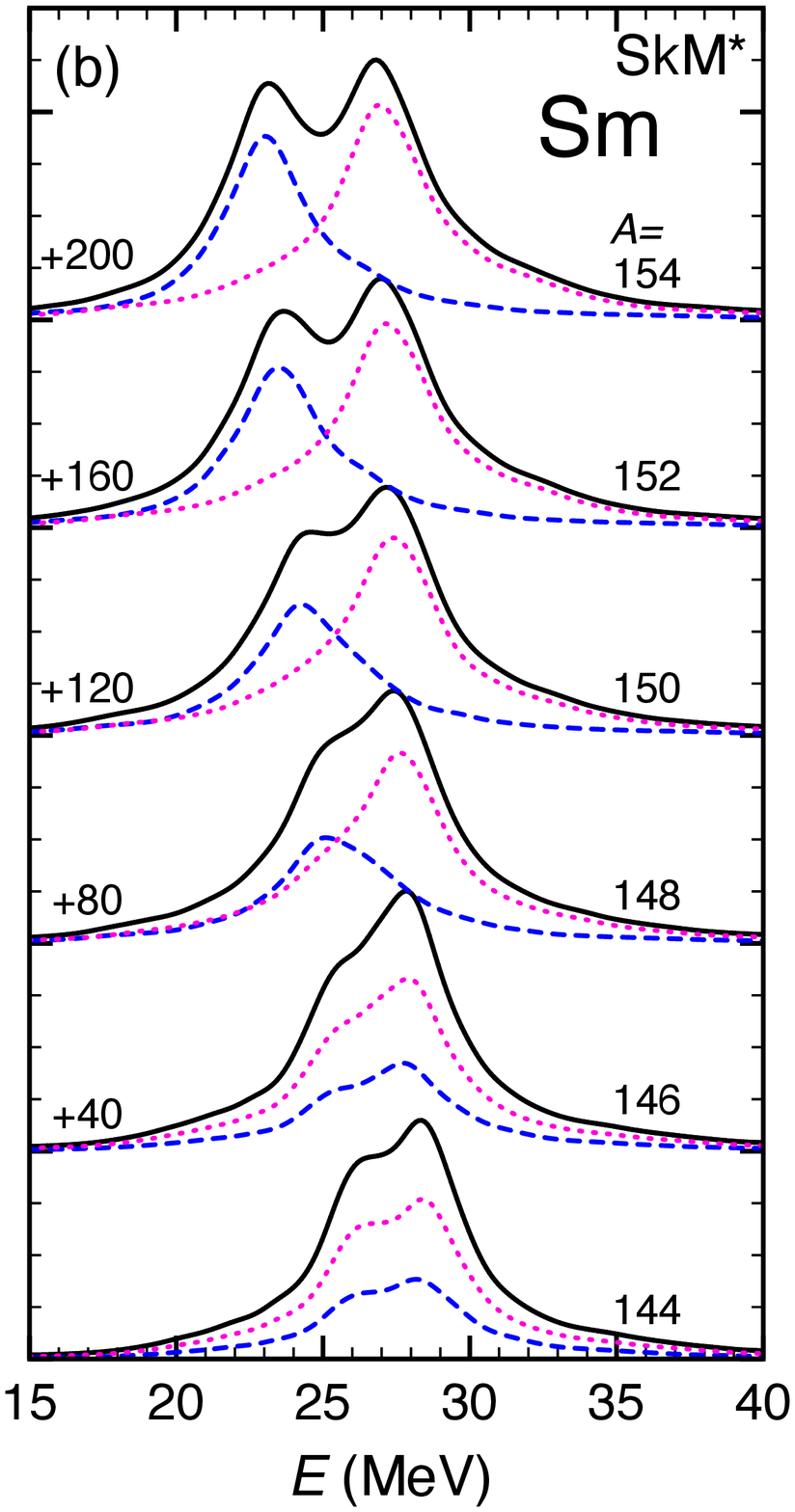}\!\!\!\!\!
\includegraphics[scale=0.32]{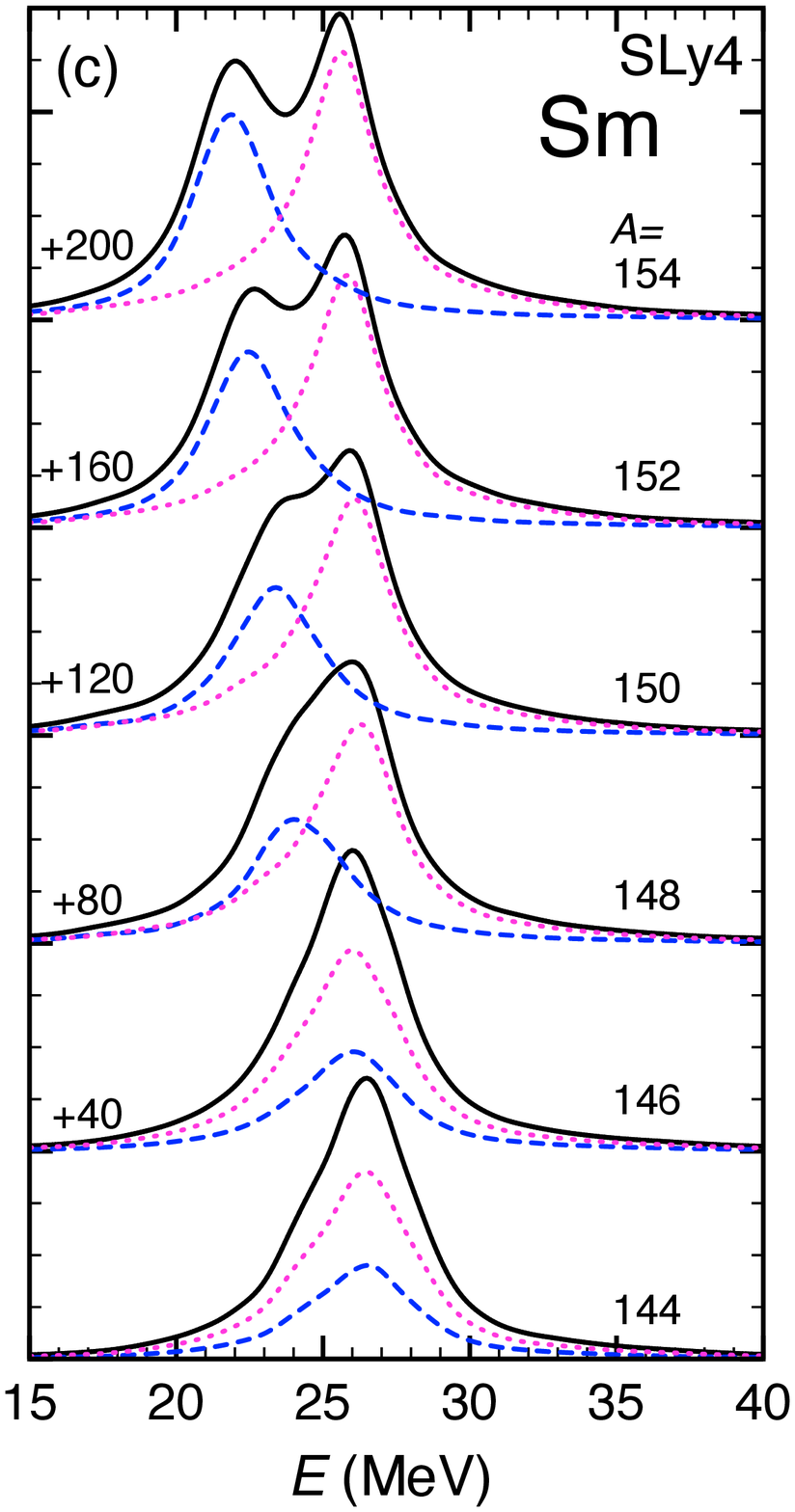}
\caption{
Calculated charge-exchange ($\tau_{-1}$) dipole strength distributions (shifted) 
in the (a) Nd  and (b) Sm isotopes with the SkM* functional, and the (c) Sm isotopes with the SLy4 functional. 
The strengths for $K=0$ and 1 excitations are drawn with the dashed and dotted line, 
and the strengths for $K=\pm1$ excitations are summed up. The smearing parameter $\gamma=2$ MeV is used. }
\label{fig:Sm}
\end{center}
\end{figure*}

Let me investigate further the roles of deformation in the charge-exchange dipole resonance.
It is observed above that a two-humped peak structure of the charge-exchange 
dipole resonance may have the same origin to that seen in the photoresonance 
in a deformed nucleus. 
It is well established that 
the photoresonance is split into two components with $K=0$ and $|K|=1$ 
in a deformed nucleus with axial symmetry, corresponding to the 
oscillations in the direction of the symmetry axis 
and those in the perpendicular directions.
The splitting is proportional to the magnitude of deformation, 
and the shape evolution has been measured in the photoabsorption cross sections 
in the rare-earth nuclei~\cite{BM2, har01}. 
Furthermore, the nuclear EDF describes well the shape change of the GDR 
in accordance with the development of nuclear deformation~\cite{kle08,yos11b, ste11, nak11, yos13b, ois16, men20}. 
One can thus expect the shape evolution to see similarly 
in the calculated charge-exchange dipole resonance.

Figure~\ref{fig:Sm} shows the transition-strength distributions for the operator $\hat{F}^-$ 
in the Nd and Sm isotopes undergoing the gradual increase in deformation. 
In this figure, the strengths for the $K=0$ and $|K|=1$ excitations are separately drawn. 
Note that the transition strengths for $K=\pm1$ excitations are summed up in plotting.  
The calculated strength distributions for $K=0$ and $K=1$ are identical to each other 
at $N=82$ and 84. 
The $K$-splitting starts to appear at $N=86$ in consonance with the appearance of deformation 
as shown in Fig.~1 of Ref.~\cite{yos11b}. 
With an increase in the neutron number, the splitting gets gradually larger 
as deformation develops. 
This is akin to the photoresonance characteristic of the rare-earth nuclei with shape evolution. 

The strength distributions in the Nd and Sm isotopes 
calculated with the use of the SkM* functional shown in Figs.~\ref{fig:Sm}(a) and \ref{fig:Sm}(b) are indicative 
of the similar nuclear structure of each isotone, 
such as the single-particle levels, unperturbed matrix elements and magnitude of deformation.
In the spherical isotones with $N=82$ and 84, one sees a shoulder structure. 
As long as the total-strength distributions are observed, 
the shoulder structure is indistinguishable from 
the $K$-splitting in the weakly-deformed nuclides with $N=86$ and 88. 
The appearance of the shoulder is also seen in the calculated photoabsorption cross sections with SkM* 
and 
it is suppressed in the calculation with the SLy4~\cite{cha98} and SkP~\cite{dob84} functionals~\cite{yos11b}. 
The detailed structure of single-particle levels affects the shape of the resonance 
through the Landau damping mechanism~\cite{ber94}. 
Figure~\ref{fig:Sm}(c) shows the strength distributions in the Sm isotopes calculated 
with the use of the SLy4 functional. 
Indeed a single peak shows up in the spherical nuclides with SLy4. 
Comparing Figs.~\ref{fig:Sm}(b) and \ref{fig:Sm}(c), one further finds that 
the peak energy calculated with SLy4 is slightly lower than that obtained with SkM* 
likewise in the calculated photoabsorption cross sections in Ref.~\cite{yos11b}.

 \begin{figure}[t]
\begin{center}
\includegraphics[scale=0.25]{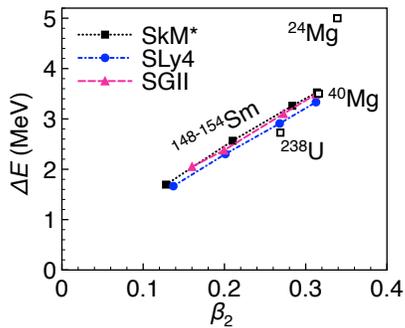}
\caption{
$K$-splitting energy for the $\Delta T_z=-1$ giant resonances in the Sm isotopes calculated by employing several Skyrme functionals. 
Included is the splitting in $^{24,40}$Mg and $^{238}$U obtained by using the SkM* functional.}
\label{fig:Ksplit}
\end{center}
\end{figure}

To see the relation between the magnitude of deformation and the evaluated splitting energy, 
I show in Fig.~\ref{fig:Ksplit} the $K$-splitting in the deformed Sm isotopes.
Here, the mean excitation energy is calculated by the moments as 
\begin{equation}
\bar{E}=\dfrac{\sum E S(E)}{\sum S(E)} 
\label{mean_e}
\end{equation}
in the energy interval of $E_1 < E< E_2$. 
Here, $E_1$ and $E_2$ are set to 15 MeV and 40 MeV. 
The change of the energy interval by a few MeV varies 
the evaluated $K$-spilling by about 0.1 -- 0.2 MeV. 
This ambiguity, however, does not affect the discussions below. 

One clearly sees a linear correlation between the magnitude of deformation, i.e. deformation parameter 
$\beta_2$, and the $K$-splitting energy, $\Delta E=\bar{E}_{K=1}-\bar{E}_{K=0}$. 
Here, the deformation parameter $\beta_2$ is defined by
\begin{equation}
\beta_2=\dfrac{4\pi}{3AR^2_{{\rm rms}}}\int d\boldsymbol{r} r^2Y_{20}(\hat{r})\varrho_0(\boldsymbol{r})
\end{equation}
with the root-mean-square radius 
$R_{\rm{rms}}=\sqrt{\frac{5}{3A}\int d\boldsymbol{r} r^2\varrho_0(\boldsymbol{r})}$ and
the isoscalar particle density $\varrho_0(\boldsymbol{r})$.
The linear correlation is also observed in the calculations employing the SLy4 and SGII~\cite{gia81} functionals.   
While the deformation property calculated with various functionals can be different, 
three lines lie close to each other. 
Note that the $K$-splitting energy calculated for a light nucleus $^{24}$Mg deviates from a trend of the Sm isotopes.

\begin{figure}[t]
\begin{center}
\includegraphics[scale=0.24]{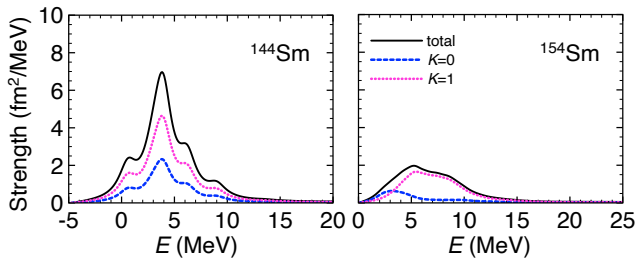}
\caption{
As Fig.~\ref{fig:Sm} but for the charge-exchange ($\tau_{+1}$) dipole operator in $^{144}$Sm and $^{154}$Sm.}
\label{fig:Sm_plus}
\end{center}
\end{figure}
 
Having the neutron excess, the strengths for $\hat{F}^+_{1K}$ are suppressed compared with those for $\hat{F}^-_{1K}$ 
as seen from Eq.~(\ref{eq:sum_rule}); 
microscopically one is due to the smaller number of proton hole states available to the dipole excitations; 
the other is due to the smaller number of neutron particle states available to the excitations, 
that is also regarded as the Pauli blocking. 
Shown in Fig.~\ref{fig:Sm_plus} is examples of the 
transition strength distributions for $\hat{F}^+_{1K}$ in the Sm isotopes. 
The strength distributions 
for $K=0$ and $|K|=1$ are identical to each other 
apart from a factor of two in the spherical nucleus $^{144}$Sm.  
In a deformed nucleus $^{154}$Sm, 
the $K$-splitting occurs as for $\hat{F}^-_{1K}$, however, 
it is difficult to see as a two-humped peak structure of the giant resonance; 
it is rather recognized as broadeneing. 
One can also see the hindrance of strengths in $^{154}$Sm than in $^{144}$Sm 
due to the neutron excess.

\begin{figure}[t]
\begin{center}
\includegraphics[scale=0.24]{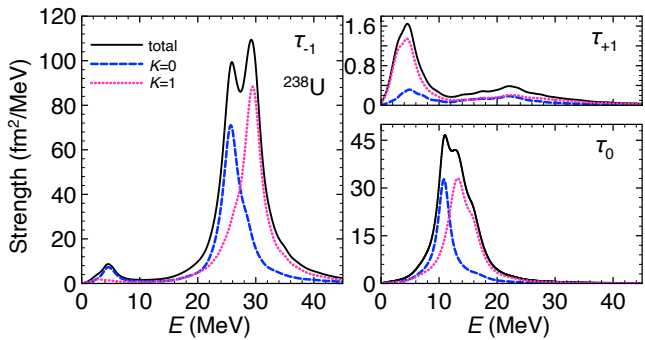}
\caption{
As Fig.~\ref{fig:24Mg} but for $^{238}$U. }
\label{fig:238U}
\end{center}
\end{figure}

The effects of neutron excess could give us a deeper understanding of the excitation modes, 
and in what follows I am going to discuss the charge-exchange dipole resonances 
in heavy nuclei and neutron-rich nuclei. 
The heavy nuclei in mid shells exhibiting a rotational spectrum are an ideal system 
to investigate the effects of deformation and neutron excess. 
I thus take the $^{238}$U nucleus as such an example, and show in Fig.~\ref{fig:238U} 
the transition-strength distributions. 
The vibrational frequency for the $\Delta T_z=\pm 1$ states is shifted relative to the $\Delta T_z=0$ state. 
In the present case, the energy difference comes from the symmetry potential associated with the neutron excess as well as 
the Coulomb energy. 
The deformation splitting of the GDR can be seen in 
the photoabsorption cross section~\cite{gur76}, 
and in the calculations~\cite{yos11,mar16a}. 
In Fig.~\ref{fig:238U}, the strength distribution for the operator $\hat{F}^0_{1K}$ is also shown for reference. 
The mean excitation energy of the 
$K=0$ and $K=1$ excitations are 11.3 MeV and 13.9 MeV, respectively. 
Thus, the $K$-splitting is evaluated as 2.6 MeV. 
Here, the mean energy was evaluated in the region of $E_1 = 5$ MeV and $E_2=35$ MeV in Eq.~(\ref{mean_e}). 
One clearly sees a deformation splitting for the charge-exchange dipole resonance for 
the operator $\hat{F}^{-}_{1K}$ similarly to the strength distributions for $\hat{F}^0_{1K}$. 
The mean excitation energy of the 
$K=0$ and $K=1$ excitations are 26.6 MeV and 29.3 MeV, respectively, 
calculated in the energy interval of 15 MeV $< E <$ 45 MeV. 
Thus, the $K$-splitting is 2.7 MeV. 
As seen in Fig.~\ref{fig:Ksplit}, 
the proportionality between the $K$-splitting and the magnitude of deformation 
lies close to the trend of the Sm isotopes. 
Though the peak energy of the $\Delta T_z=-1$ state is higher than that of the $\Delta T_z=0$ state, 
one finds the same amount of $K$-splitting for the giant resonance. 
One cannot see, however, the $K$-splitting in the $\Delta T_z=+1$ giant resonance. 
Though the transition strengths are gathered in low energy,  
they are quite hindered compared to those of the 
$\Delta T_z=0$ and $\Delta T_z=-1$ giant resonances, indicating a weak collectivity. 

One notices the appearance of the dipole states in low energy $E < 10$ MeV 
for the response to the operator $\hat{F}^{-}_{1K}$. 
The low-energy dipole states correspond to the $-1\hbar \omega_0$ excitation~\cite{yos17} 
uniquely appearing in nuclei with neutron excess.  
The excitation of this type is associated with the Fermi levels 
of protons and neutrons being apart by one major shell. 
The lowest energy particle-hole or 2qp excitations are thus negative parity. 
In the present case, the Nilsson orbitals stemming from the $j_{15/2}$ shell 
are partly occupied by neutrons while those from the $i_{13/2}$ shell are almost empty 
for protons. 
Since the number of 2qp excitations 
satisfying the selection rule for the transition is not large as seen in Ref.~\cite{yos17}, 
the collectivity is weak and the excitation energy is sensitive to the details of 
shell structure around the Fermi levels. 

\begin{figure}[t]
\begin{center}
\includegraphics[scale=0.24]{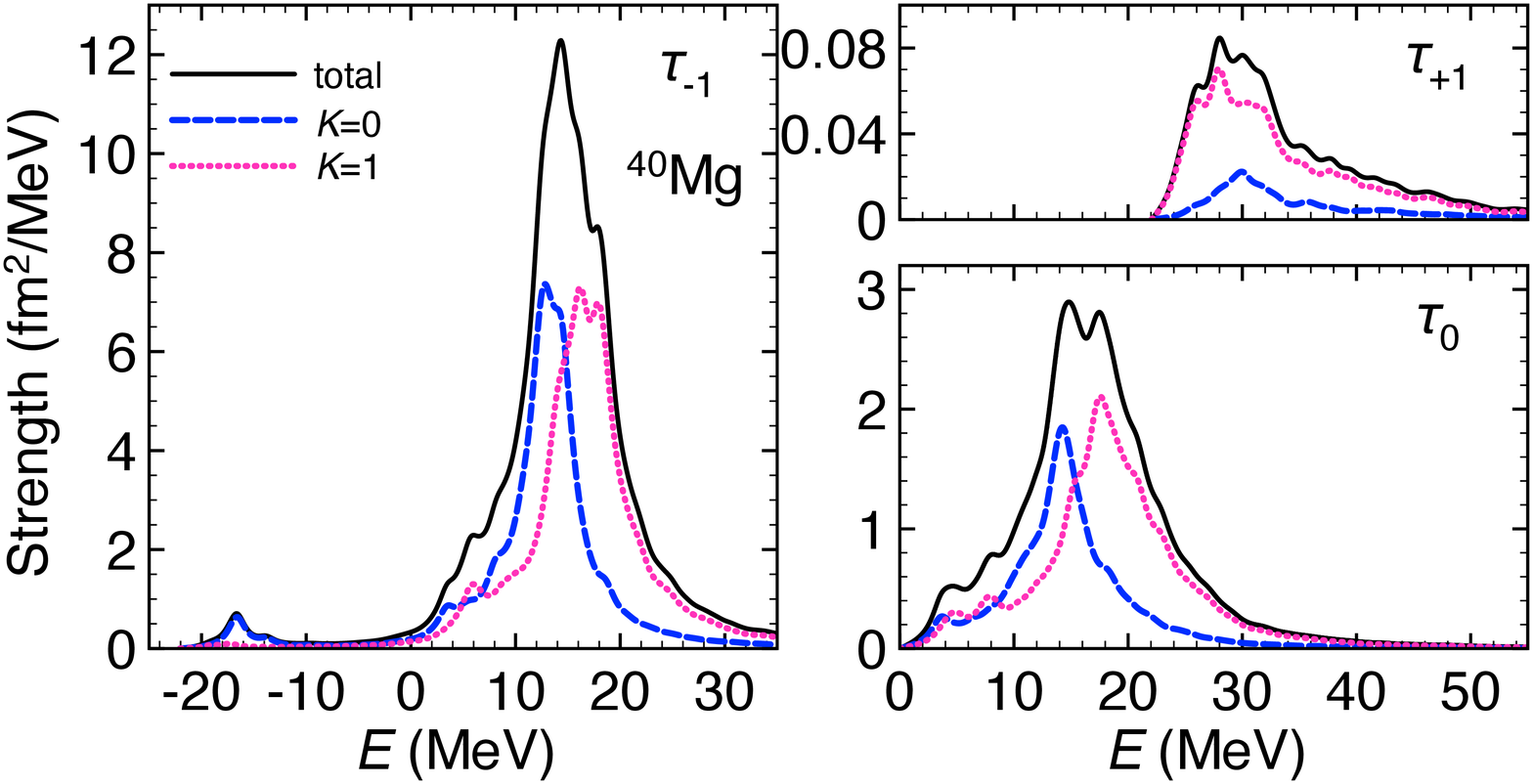}
\caption{
As Fig.~\ref{fig:24Mg} but for $^{40}$Mg.}
\label{fig:40Mg}
\end{center}
\end{figure}

At the end of investigation of the dipole resonances in nuclei with neutron excess, 
I discuss the strength distributions in neutron-rich exotic nuclei. 
The $^{40}$Mg nucleus has attracted interest in a possible quadrupole deformation 
due to the broken spherical magic number of $N=28$ near the drip line~\cite{doo13,cra19}.
Theoretically, the deformation properties of the Mg isotopes close to the drip line 
have been explored by the Skyrme~\cite{ter97,sto03,yos09a,pei09,yam19}, Gogny~\cite{rod02} and relativistic~\cite{li12} EDF approaches, 
and the GDR as well as the LED/PDR are predicted by the Skyrme EDF calculations~\cite{yos09,wan17b}.
Thus, 
I investigate here the charge-exchange dipole responses in $^{40}$Mg to see 
the effects of deformation and an extreme neutron excess. 

Figure~\ref{fig:40Mg} displays the strength distributions in $^{40}$Mg. 
The $K$-splitting associated with deformation emerges for the giant resonances. 
The mean energy of the $K=0$ and $K=1$ excitations of the giant resonance 
is 13.2 MeV and 16.7 MeV, respectively. 
Here, the energy interval is set as $0$ MeV  $< E < 40$ MeV 
though this may include the effect of LED. 
The calculated $K$-splitting energy for the $\Delta T_z=-1$ giant resonance 
3.5 MeV is compatible with 3.8 MeV for the $\Delta T_z=0$ giant resonance. 
One of the common features in nuclei with neutron excess is that 
the strengths for the $\Delta T_z=-1$ excitation are enhanced while those for the $\Delta T_z=+1$ excitation are suppressed. 
Furthermore, one sees occurrence of the $-1\hbar \omega_0$ excitation. 
In the present case, the Nilsson orbitals stemming from the $p_{3/2}$ and $f_{7/2}$ shells are 
mostly occupied by neutrons while the $sd$ shell is almost empty for protons. 
This is an ideal situation where the negative-parity excitations appear in low energy.

One sees, however, some distinct features in neutron-rich unstable nuclei. 
The ordering of the excitation energies of the $\Delta T_z=\pm 1$ and $0$ giant resonances are different 
from those in the stable nuclei observed so far. 
In the present case, one sees $E_{\Delta T_z=0}\simeq E_{\Delta T_z=-1}<E_{\Delta T_z=+1}$, 
while one found 
$E_{\Delta T_z=+1}< E_{\Delta T_z=0}<E_{\Delta T_z=-1}$ in stable nuclei. 
This anomalous behavior is due to the highly imbalanced Fermi levels of protons and neutrons. 
As one sees in the figure, the strengths appear around $ -20$ MeV for the $\Delta T_z=-1$ excitation, 
and show up only above $\sim 20$ MeV for the $\Delta T_z=+1$ excitation. 
Furthermore, the concentration of transition strengths and a shoulder structure emerge below the giant resonance. 
The concentration of transition strengths below the giant resonance for the $\Delta T_z=-1$ excitation 
is constructed by the 2qp excitations involving the weakly-bound and quasiparticle-resonant neutrons near 
the threshold such as in the $f_{7/2}$ shell
and the proton continuum states in the $g_{9/2}$ shell. 
Since the LED states below $10$ MeV 
for the $\Delta T_z=0$ excitation are generated by the continuum states 
of neutrons in the $g_{9/2}$ shell ~\cite{yos09} instead of protons, 
the shoulder structure below $E\sim 10$ MeV for the $\Delta T_z=-1$ excitation 
can be considered as an analog of the LED and a unique feature of the charge-exchange dipole resonance in 
neutron-rich unstable nuclei 
as previously discussed in spherical nuclei~\cite{yos17}.

\subsection{Spin dipole excitations}\label{result_SD}

\begin{figure*}[t]
\begin{center}
\includegraphics[scale=0.25]{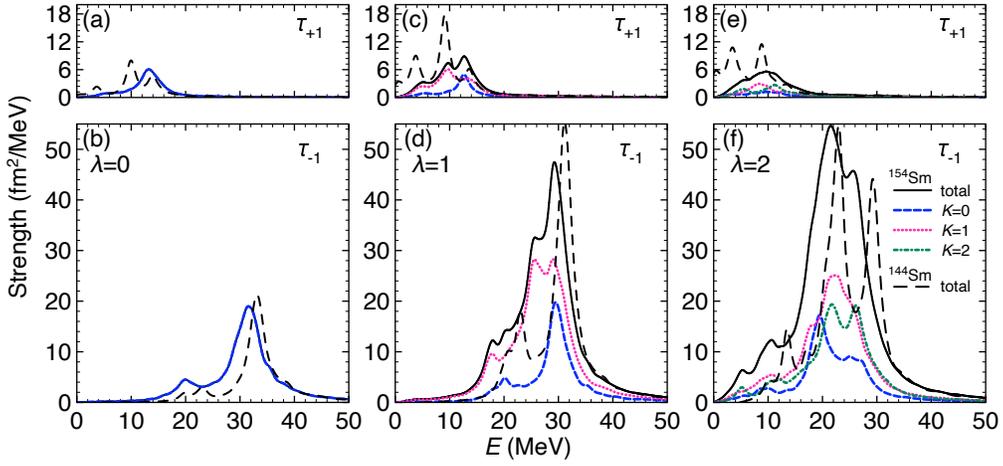}
\caption{
Calculated spin-dipole strength distributions in $^{144}$Sm and $^{154}$Sm obtained by using the SkM* functional.
}
\label{fig:SD}
\end{center}
\end{figure*}

Let me investigate briefly the deformation effects on the spin-dipole (SD) strengths. 
I show in Fig.~\ref{fig:SD} the strength distributions in $^{144}$Sm and $^{154}$Sm. 
Here, the charge-exchange rank-$\lambda$ SD operators are defied as
\begin{equation}
\hat{F}^\pm_{\lambda K} = \sum_{\sigma, \sigma^\prime, \tau, \tau^\prime}\int d\boldsymbol{r} 
r[Y_1\otimes \boldsymbol{\sigma}]^\lambda_K \langle \tau |\tau_{\pm 1}|\tau^\prime\rangle
\hat{\psi}^\dagger(\boldsymbol{r}  \sigma \tau)  \hat{\psi}(\boldsymbol{r}  \sigma^\prime \tau^\prime), \label{sd_op}
\end{equation}
where 
$[Y_1\otimes \boldsymbol{\sigma}]^\lambda_K=\sum_{\mu \nu}\langle 1\mu1\nu|\lambda K\rangle Y_{1\mu}
\langle\sigma|\sigma_\nu|\sigma^\prime\rangle$
with the spherical components of the Pauli spin matrix $\vec{\sigma}=(\sigma_{-1},\sigma_0,\sigma_{+1})$. 
Since there is no $K$-dependence in spherical nuclei, the strengths for each $K$-component are summed up 
in drawing the strength distribution for $^{144}$Sm. 
The strengths of $\pm K$ are summed up as above.
The strengths in the $\tau_+$ channel [Figs.~\ref{fig:SD}(a), \ref{fig:SD}(c), \ref{fig:SD}(e)] are hindered compared 
with those in the $\tau_-$ channel [Figs.~\ref{fig:SD}(b), \ref{fig:SD}(d), \ref{fig:SD}(f)] 
as in the case for the electric (non-spin-flip) dipole resonance due to the neutron excess. 
The deformation effects can be observed as broadening of the resonance. 
Let me discuss below the strengths in the $\tau_-$ channel. 

Irrespective of deformation, the transition strengths are scattered into three components 
of $\lambda=0, 1$ and $2$. 
The mean energy is 30.4 (33.3) MeV, 28.8 (30.0) MeV and 21.4 (24.7) MeV for the $\lambda=0, 1$ and 2 components, respectively,  
in $^{154}$Sm ($^{144}$Sm). 
Here, the energy interval is set as 0 MeV $< E<$ 50 MeV. The energies calculated here including the even-$N$ $^{146 \textendash 152}$Sm follow the systematic trend 
with $\lambda=0$ being highest and $\lambda=2$ lowest~\cite{ber81}.

To investigate the strength distributions, it may be helpful to see the model-independent sum rules 
for the SD operators~\cite{aue84}. 
The sum rules generalized to the deformed systems read
\begin{align}
&\int \dd E [S^-_{\lambda K}(E)-S^+_{\lambda K}(E)] \notag \\
&=
\left\{
\begin{aligned}
2\dfrac{1}{4\pi}[N\langle r^2 \rangle_N - Z\langle r^2 \rangle_Z]& \,\,\, (\lambda=0, K=0) \\
2\dfrac{3}{8\pi}[N\langle \rho^2 \rangle_N - Z\langle \rho^2 \rangle_Z]& \,\,\, (\lambda=1, K=0) \\
2\dfrac{3}{16\pi}[N\langle \rho^2+2z^2 \rangle_N - Z\langle \rho^2+2z^2\rangle_Z]& \,\,\, (\lambda=1, K=\pm1) \\
2\dfrac{1}{8\pi}[N\langle \rho^2+4z^2 \rangle_N - Z\langle \rho^2+4z^2\rangle_Z]& \,\,\, (\lambda=2, K=0) \\
2\dfrac{3}{16\pi}[N\langle \rho^2+2z^2 \rangle_N - Z\langle \rho^2+2z^2\rangle_Z]& \,\,\, (\lambda=2, K=	\pm1) \\
2\dfrac{3}{8\pi}[N\langle \rho^2 \rangle_N - Z\langle \rho^2\rangle_Z]& \,\,\, (\lambda=2, K=\pm2)
\end{aligned}
\right.,
\label{eq:SD_sum_rule}
\end{align}
and coincide with $\dfrac{1}{2\pi}[N\langle r^2\rangle_N - Z\langle r^2\rangle_Z]$ in the spherical limit for each $K$-component
as given in Ref.~\cite{aue84}.  In deriving the formulae, the time-reversal symmetry of the ground state was assumed. 
 
The strength distribution for the rank-$0$ SD operator is shown in Fig.~\ref{fig:SD}(b). 
Since the rank-$0$ operator is scalar, one has no $K$-dependence even in deformed nuclei. 
Indeed, the line shapes for $^{144}$Sm and $^{154}$Sm 
are similar to each other. 
One sees $K$-splitting in the strength distributions for the rank-$1$ and $2$ operators as shown in Figs.~\ref{fig:SD}(d) and \ref{fig:SD}(f). 
The simple geometrical argument for the $K$-splitting in the electric dipole resonance cannot be applied. 
In the case of the non-spin-flip dipole resonance, the $K$-dependence comes from 
the spherical harmonics $Y_{1K}$ representing the nuclear shape in real space. 
When the spin degree of freedom is involved in the case of the SD resonance, 
the $K$ quantum number does not directly characterize the nuclear deformation represented by $\mu$
in the definition of the operator Eq.~(\ref{sd_op}). 
However, a qualitative argument on the $K$-dependence of the strengths can be given according to the sum rules. 
To first order in deformation, one can express $\langle z^2 \rangle=\frac{1}{3}\langle r^2\rangle(1+\frac{2}{3}\delta)$ and 
 $\langle \rho^2 \rangle=\frac{2}{3}\langle r^2\rangle(1-\frac{1}{3}\delta)$ with $\delta$ representing 
the deformation parameter~\cite{BM2}.  
The $K=0$ strength is reduced by $\frac{1}{6\pi}(N\langle r^2 \rangle_N- Z\langle r^2 \rangle_Z) \delta$ while 
the $K=1$ strength is enhanced by $\frac{1}{12\pi}(N\langle r^2 \rangle_N- Z\langle r^2 \rangle_Z) \delta$ for $\lambda=1$. 
For $\lambda=2$, 
the $K=0$ and $K=1$ strengths are enhanced by  $\frac{1}{6\pi}(N\langle r^2 \rangle_N- Z\langle r^2 \rangle_Z)\delta$ and 
$\frac{1}{12\pi}(N\langle r^2 \rangle_N- Z\langle r^2 \rangle_Z) \delta$, respectively 
and the $K=2$ strength is reduced by $\frac{1}{6\pi}(N\langle r^2 \rangle_N- Z\langle r^2 \rangle_Z) \delta$. 
The summed strengths for each $\lambda$ are unchanged within this approximation. 

One sees a two-peak structure of the SDR in $^{144}$Sm 
and a broad resonance structure in $^{154}$Sm for the $\lambda=1$ excitation.
As expected from the sum rules, the $K=0$ strength is reduced while the $K=1$ strength is enhanced due to deformation. 
Broadening of the resonance can be seen also for the $\lambda=2$ excitation. 
One observes that the split states are overlapping. 
It is clearly seen that the $K=2$ strength decreases. 
The precedent nuclear EDF calculations~\cite{suz00,sag07,fra07,lia12}, though restricted to spherical nuclei, 
predict that the fragmentation increases with $\lambda$. 
Following the early findings, one sees that the strengths for $\lambda=2$ are fragmented in $^{144}$Sm. 
Therefore, the deformation-induced broadening is unlikely to observe experimentally 
if the spreading width is $\gtrsim$ 2 MeV. 
Furthermore, 
desired are further attempts of 
disentangling the multipolarity $\lambda$ and reducing the continuum background 
to extract the details of resonance structure~\cite{aki94,oka95,deh07,doz08,doz20}. 

\section{Summary}\label{summary}

The deformation effects on the charge-exchange electric (non-spin-flip) and magnetic (spin-flip) 
dipole excitations were investigated 
by means of the fully self-consistent pnQRPA with the Skyrme EDF. 
I found that 
the deformation splitting into $K=0$ and $K=\pm 1$ components occurs generically 
for the IV electric dipole resonance and is proportional to 
the magnitude of deformation. 
The $K$-splitting shows up also for the charge-exchange magnetic dipole resonance. 
However, a simple geometrical assertion valid for the electric cases 
cannot be applied for explaining the vibrational frequencies of each $K$-component  
due to the coupling of spin and angular momentum in the magnetic excitations.  
The model-independent non-energy-weighted sum rules were derived for the axially-deformed nuclei, 
and a qualitative argument on the structure of strength distributions for each $K$-component 
was given. 
The IVGDR and IV low-energy octupole resonance can couple in deformed nuclei, 
and the shoulder structure in the octupole resonance is predicted to appear 
due to these coupling~\cite{yos13b}. 
It is thus an interesting future study to see if the coupling effects show up generally in the charge-exchange 
octupole resonances in deformed nuclei.  
In nuclei with an appreciable neutron excess, I found the concentration of the dipole strengths in low energy 
and a shoulder structure below the giant resonance. 
These modes of excitation are unique in neutron-rich unstable nuclei 
and can emerge in deformed nuclei as well as in spherical systems~\cite{yos17}. 

\begin{acknowledgments} 
This work was supported by the JSPS KAKENHI (Grant Nos. JP18H04569, JP19K03824 and JP19K03872).
The numerical calculations were performed on CRAY XC40 
at the Yukawa Institute for Theoretical Physics, Kyoto University.
\end{acknowledgments}

\bibliography{Def_dipole_ref}

\begin{thebibliography}{75}%
\makeatletter
\providecommand \@ifxundefined [1]{%
 \@ifx{#1\undefined}
}%
\providecommand \@ifnum [1]{%
 \ifnum #1\expandafter \@firstoftwo
 \else \expandafter \@secondoftwo
 \fi
}%
\providecommand \@ifx [1]{%
 \ifx #1\expandafter \@firstoftwo
 \else \expandafter \@secondoftwo
 \fi
}%
\providecommand \natexlab [1]{#1}%
\providecommand \enquote  [1]{``#1''}%
\providecommand \bibnamefont  [1]{#1}%
\providecommand \bibfnamefont [1]{#1}%
\providecommand \citenamefont [1]{#1}%
\providecommand \href@noop [0]{\@secondoftwo}%
\providecommand \href [0]{\begingroup \@sanitize@url \@href}%
\providecommand \@href[1]{\@@startlink{#1}\@@href}%
\providecommand \@@href[1]{\endgroup#1\@@endlink}%
\providecommand \@sanitize@url [0]{\catcode `\\12\catcode `\$12\catcode
  `\&12\catcode `\#12\catcode `\^12\catcode `\_12\catcode `\%12\relax}%
\providecommand \@@startlink[1]{}%
\providecommand \@@endlink[0]{}%
\providecommand \url  [0]{\begingroup\@sanitize@url \@url }%
\providecommand \@url [1]{\endgroup\@href {#1}{\urlprefix }}%
\providecommand \urlprefix  [0]{URL }%
\providecommand \Eprint [0]{\href }%
\providecommand \doibase [0]{http://dx.doi.org/}%
\providecommand \selectlanguage [0]{\@gobble}%
\providecommand \bibinfo  [0]{\@secondoftwo}%
\providecommand \bibfield  [0]{\@secondoftwo}%
\providecommand \translation [1]{[#1]}%
\providecommand \BibitemOpen [0]{}%
\providecommand \bibitemStop [0]{}%
\providecommand \bibitemNoStop [0]{.\EOS\space}%
\providecommand \EOS [0]{\spacefactor3000\relax}%
\providecommand \BibitemShut  [1]{\csname bibitem#1\endcsname}%
\let\auto@bib@innerbib\@empty
\bibitem [{\citenamefont {Osterfeld}(1992)}]{ost92}%
  \BibitemOpen
  \bibfield  {author} {\bibinfo {author} {\bibfnamefont {F.}~\bibnamefont
  {Osterfeld}},\ }\href {\doibase 10.1103/RevModPhys.64.491} {\bibfield
  {journal} {\bibinfo  {journal} {Rev. Mod. Phys.}\ }\textbf {\bibinfo {volume}
  {64}},\ \bibinfo {pages} {491} (\bibinfo {year} {1992})}\BibitemShut
  {NoStop}%
\bibitem [{\citenamefont {Ejiri}(2000)}]{eji00}%
  \BibitemOpen
  \bibfield  {author} {\bibinfo {author} {\bibfnamefont {H.}~\bibnamefont
  {Ejiri}},\ }\href {\doibase https://doi.org/10.1016/S0370-1573(00)00044-2}
  {\bibfield  {journal} {\bibinfo  {journal} {Phys. Rep.}\ }\textbf {\bibinfo
  {volume} {338}},\ \bibinfo {pages} {265 } (\bibinfo {year}
  {2000})}\BibitemShut {NoStop}%
\bibitem [{\citenamefont {Measday}(2001)}]{mea01}%
  \BibitemOpen
  \bibfield  {author} {\bibinfo {author} {\bibfnamefont {D.}~\bibnamefont
  {Measday}},\ }\href {\doibase https://doi.org/10.1016/S0370-1573(01)00012-6}
  {\bibfield  {journal} {\bibinfo  {journal} {Phys. Rep.}\ }\textbf {\bibinfo
  {volume} {354}},\ \bibinfo {pages} {243 } (\bibinfo {year}
  {2001})}\BibitemShut {NoStop}%
\bibitem [{\citenamefont {Langanke}\ and\ \citenamefont
  {Mart\'inez-Pinedo}(2003)}]{lan03}%
  \BibitemOpen
  \bibfield  {author} {\bibinfo {author} {\bibfnamefont {K.}~\bibnamefont
  {Langanke}}\ and\ \bibinfo {author} {\bibfnamefont {G.}~\bibnamefont
  {Mart\'inez-Pinedo}},\ }\href {\doibase 10.1103/RevModPhys.75.819} {\bibfield
   {journal} {\bibinfo  {journal} {Rev. Mod. Phys.}\ }\textbf {\bibinfo
  {volume} {75}},\ \bibinfo {pages} {819} (\bibinfo {year} {2003})},\ \Eprint
  {http://arxiv.org/abs/nucl-th/0203071} {arXiv:nucl-th/0203071} \BibitemShut
  {NoStop}%
\bibitem [{\citenamefont {Ichimura}\ \emph {et~al.}(2006)\citenamefont
  {Ichimura}, \citenamefont {Sakai},\ and\ \citenamefont {Wakasa}}]{ich06}%
  \BibitemOpen
  \bibfield  {author} {\bibinfo {author} {\bibfnamefont {M.}~\bibnamefont
  {Ichimura}}, \bibinfo {author} {\bibfnamefont {H.}~\bibnamefont {Sakai}}, \
  and\ \bibinfo {author} {\bibfnamefont {T.}~\bibnamefont {Wakasa}},\ }\href
  {\doibase 10.1016/j.ppnp.2005.09.001} {\bibfield  {journal} {\bibinfo
  {journal} {Prog. Part. Nucl. Phys.}\ }\textbf {\bibinfo {volume} {56}},\
  \bibinfo {pages} {446} (\bibinfo {year} {2006})}\BibitemShut {NoStop}%
\bibitem [{\citenamefont {Avignone}\ \emph {et~al.}(2008)\citenamefont
  {Avignone}, \citenamefont {Elliott},\ and\ \citenamefont {Engel}}]{avi08}%
  \BibitemOpen
  \bibfield  {author} {\bibinfo {author} {\bibfnamefont {F.~T.}\ \bibnamefont
  {Avignone}}, \bibinfo {author} {\bibfnamefont {S.~R.}\ \bibnamefont
  {Elliott}}, \ and\ \bibinfo {author} {\bibfnamefont {J.}~\bibnamefont
  {Engel}},\ }\href {\doibase 10.1103/RevModPhys.80.481} {\bibfield  {journal}
  {\bibinfo  {journal} {Rev. Mod. Phys.}\ }\textbf {\bibinfo {volume} {80}},\
  \bibinfo {pages} {481} (\bibinfo {year} {2008})},\ \Eprint
  {http://arxiv.org/abs/0708.1033} {arXiv:0708.1033 [nucl-ex]} \BibitemShut
  {NoStop}%
\bibitem [{\citenamefont {Fujita}\ \emph {et~al.}(2011)\citenamefont {Fujita},
  \citenamefont {Rubio},\ and\ \citenamefont {Gelletly}}]{fuj11}%
  \BibitemOpen
  \bibfield  {author} {\bibinfo {author} {\bibfnamefont {Y.}~\bibnamefont
  {Fujita}}, \bibinfo {author} {\bibfnamefont {B.}~\bibnamefont {Rubio}}, \
  and\ \bibinfo {author} {\bibfnamefont {W.}~\bibnamefont {Gelletly}},\ }\href
  {\doibase https://doi.org/10.1016/j.ppnp.2011.01.056} {\bibfield  {journal}
  {\bibinfo  {journal} {Prog. Part. Nucl. Phys.}\ }\textbf {\bibinfo {volume}
  {66}},\ \bibinfo {pages} {549 } (\bibinfo {year} {2011})}\BibitemShut
  {NoStop}%
\bibitem [{\citenamefont {Nakamura}\ \emph {et~al.}(2017)\citenamefont
  {Nakamura}, \citenamefont {Kamano}, \citenamefont {Hayato}, \citenamefont
  {Hirai}, \citenamefont {Horiuchi}, \citenamefont {Kumano}, \citenamefont
  {Murata}, \citenamefont {Saito}, \citenamefont {Sakuda}, \citenamefont
  {Sato},\ and\ \citenamefont {Suzuki}}]{nak17}%
  \BibitemOpen
  \bibfield  {author} {\bibinfo {author} {\bibfnamefont {S.~X.}\ \bibnamefont
  {Nakamura}}, \bibinfo {author} {\bibfnamefont {H.}~\bibnamefont {Kamano}},
  \bibinfo {author} {\bibfnamefont {Y.}~\bibnamefont {Hayato}}, \bibinfo
  {author} {\bibfnamefont {M.}~\bibnamefont {Hirai}}, \bibinfo {author}
  {\bibfnamefont {W.}~\bibnamefont {Horiuchi}}, \bibinfo {author}
  {\bibfnamefont {S.}~\bibnamefont {Kumano}}, \bibinfo {author} {\bibfnamefont
  {T.}~\bibnamefont {Murata}}, \bibinfo {author} {\bibfnamefont
  {K.}~\bibnamefont {Saito}}, \bibinfo {author} {\bibfnamefont
  {M.}~\bibnamefont {Sakuda}}, \bibinfo {author} {\bibfnamefont
  {T.}~\bibnamefont {Sato}}, \ and\ \bibinfo {author} {\bibfnamefont
  {Y.}~\bibnamefont {Suzuki}},\ }\href {\doibase 10.1088/1361-6633/aa5e6c}
  {\bibfield  {journal} {\bibinfo  {journal} {Rept. Prog. Phys.}\ }\textbf
  {\bibinfo {volume} {80}},\ \bibinfo {pages} {056301} (\bibinfo {year}
  {2017})},\ \Eprint {http://arxiv.org/abs/1610.01464} {arXiv:1610.01464
  [nucl-th]} \BibitemShut {NoStop}%
\bibitem [{\citenamefont {Lenske}\ \emph {et~al.}(2019)\citenamefont {Lenske},
  \citenamefont {Cappuzzello}, \citenamefont {Cavallaro},\ and\ \citenamefont
  {Colonna}}]{len19}%
  \BibitemOpen
  \bibfield  {author} {\bibinfo {author} {\bibfnamefont {H.}~\bibnamefont
  {Lenske}}, \bibinfo {author} {\bibfnamefont {F.}~\bibnamefont {Cappuzzello}},
  \bibinfo {author} {\bibfnamefont {M.}~\bibnamefont {Cavallaro}}, \ and\
  \bibinfo {author} {\bibfnamefont {M.}~\bibnamefont {Colonna}},\ }\href
  {\doibase 10.1016/j.ppnp.2019.103716} {\bibfield  {journal} {\bibinfo
  {journal} {Prog. Part. Nucl. Phys.}\ }\textbf {\bibinfo {volume} {109}},\
  \bibinfo {pages} {103716} (\bibinfo {year} {2019})}\BibitemShut {NoStop}%
\bibitem [{\citenamefont {Harakeh}\ and\ \citenamefont {Woude}(2001)}]{har01}%
  \BibitemOpen
  \bibfield  {author} {\bibinfo {author} {\bibfnamefont {M.}~\bibnamefont
  {Harakeh}}\ and\ \bibinfo {author} {\bibfnamefont {A.}~\bibnamefont
  {Woude}},\ }\href {https://books.google.co.jp/books?id=ux0JhIdbGT8C} {\emph
  {\bibinfo {title} {Giant Resonances: Fundamental High-frequency Modes of
  Nuclear Excitation}}},\ Oxford science publications\ (\bibinfo  {publisher}
  {Oxford University Press},\ \bibinfo {year} {2001})\BibitemShut {NoStop}%
\bibitem [{\citenamefont {Berman}\ and\ \citenamefont {Fultz}(1975)}]{ber75}%
  \BibitemOpen
  \bibfield  {author} {\bibinfo {author} {\bibfnamefont {B.~L.}\ \bibnamefont
  {Berman}}\ and\ \bibinfo {author} {\bibfnamefont {S.~C.}\ \bibnamefont
  {Fultz}},\ }\href {\doibase 10.1103/RevModPhys.47.713} {\bibfield  {journal}
  {\bibinfo  {journal} {Rev. Mod. Phys.}\ }\textbf {\bibinfo {volume} {47}},\
  \bibinfo {pages} {713} (\bibinfo {year} {1975})}\BibitemShut {NoStop}%
\bibitem [{\citenamefont {Paar}\ \emph {et~al.}(2007)\citenamefont {Paar},
  \citenamefont {Vretenar}, \citenamefont {Khan},\ and\ \citenamefont
  {Col\`o}}]{par07}%
  \BibitemOpen
  \bibfield  {author} {\bibinfo {author} {\bibfnamefont {N.}~\bibnamefont
  {Paar}}, \bibinfo {author} {\bibfnamefont {D.}~\bibnamefont {Vretenar}},
  \bibinfo {author} {\bibfnamefont {E.}~\bibnamefont {Khan}}, \ and\ \bibinfo
  {author} {\bibfnamefont {G.}~\bibnamefont {Col\`o}},\ }\href {\doibase
  10.1088/0034-4885/70/5/R02} {\bibfield  {journal} {\bibinfo  {journal} {Rept.
  Prog. Phys.}\ }\textbf {\bibinfo {volume} {70}},\ \bibinfo {pages} {691}
  (\bibinfo {year} {2007})},\ \Eprint {http://arxiv.org/abs/nucl-th/0701081}
  {arXiv:nucl-th/0701081} \BibitemShut {NoStop}%
\bibitem [{\citenamefont {Aumann}\ and\ \citenamefont
  {Nakamura}(2013)}]{aum13}%
  \BibitemOpen
  \bibfield  {author} {\bibinfo {author} {\bibfnamefont {T.}~\bibnamefont
  {Aumann}}\ and\ \bibinfo {author} {\bibfnamefont {T.}~\bibnamefont
  {Nakamura}},\ }\href {\doibase 10.1088/0031-8949/2013/t152/014012} {\bibfield
   {journal} {\bibinfo  {journal} {Phys. Scr.}\ }\textbf {\bibinfo {volume}
  {T152}},\ \bibinfo {pages} {014012} (\bibinfo {year} {2013})}\BibitemShut
  {NoStop}%
\bibitem [{\citenamefont {Roca-Maza}\ and\ \citenamefont {Paar}(2018)}]{roc18}%
  \BibitemOpen
  \bibfield  {author} {\bibinfo {author} {\bibfnamefont {X.}~\bibnamefont
  {Roca-Maza}}\ and\ \bibinfo {author} {\bibfnamefont {N.}~\bibnamefont
  {Paar}},\ }\href {\doibase 10.1016/j.ppnp.2018.04.001} {\bibfield  {journal}
  {\bibinfo  {journal} {Prog. Part. Nucl. Phys.}\ }\textbf {\bibinfo {volume}
  {101}},\ \bibinfo {pages} {96} (\bibinfo {year} {2018})},\ \Eprint
  {http://arxiv.org/abs/1804.06256} {arXiv:1804.06256 [nucl-th]} \BibitemShut
  {NoStop}%
\bibitem [{\citenamefont {Bracco}\ \emph {et~al.}(2019)\citenamefont {Bracco},
  \citenamefont {Lanza},\ and\ \citenamefont {Tamii}}]{bra19}%
  \BibitemOpen
  \bibfield  {author} {\bibinfo {author} {\bibfnamefont {A.}~\bibnamefont
  {Bracco}}, \bibinfo {author} {\bibfnamefont {E.}~\bibnamefont {Lanza}}, \
  and\ \bibinfo {author} {\bibfnamefont {A.}~\bibnamefont {Tamii}},\ }\href
  {\doibase https://doi.org/10.1016/j.ppnp.2019.02.001} {\bibfield  {journal}
  {\bibinfo  {journal} {Prog. Part. Nucl. Phys.}\ }\textbf {\bibinfo {volume}
  {106}},\ \bibinfo {pages} {360 } (\bibinfo {year} {2019})}\BibitemShut
  {NoStop}%
\bibitem [{\citenamefont {Bohr}\ and\ \citenamefont {Mottelson}(1969)}]{BM2}%
  \BibitemOpen
  \bibfield  {author} {\bibinfo {author} {\bibfnamefont {A.}~\bibnamefont
  {Bohr}}\ and\ \bibinfo {author} {\bibfnamefont {B.}~\bibnamefont
  {Mottelson}},\ }\href {https://books.google.co.jp/books?id=dRhRAAAAMAAJ}
  {\emph {\bibinfo {title} {Nuclear Structure: Volume II, Nuclear
  Deformations}}}\ (\bibinfo  {publisher} {Benjamin},\ \bibinfo {year}
  {1969})\BibitemShut {NoStop}%
\bibitem [{\citenamefont {Izumoto}(1983)}]{izu83}%
  \BibitemOpen
  \bibfield  {author} {\bibinfo {author} {\bibfnamefont {T.}~\bibnamefont
  {Izumoto}},\ }\href {\doibase https://doi.org/10.1016/0375-9474(83)90095-7}
  {\bibfield  {journal} {\bibinfo  {journal} {Nucl. Phys. A}\ }\textbf
  {\bibinfo {volume} {395}},\ \bibinfo {pages} {189 } (\bibinfo {year}
  {1983})}\BibitemShut {NoStop}%
\bibitem [{\citenamefont {Auerbach}\ and\ \citenamefont {Klein}(1984)}]{aue84}%
  \BibitemOpen
  \bibfield  {author} {\bibinfo {author} {\bibfnamefont {N.}~\bibnamefont
  {Auerbach}}\ and\ \bibinfo {author} {\bibfnamefont {A.}~\bibnamefont
  {Klein}},\ }\href {\doibase 10.1103/PhysRevC.30.1032} {\bibfield  {journal}
  {\bibinfo  {journal} {Phys. Rev. C}\ }\textbf {\bibinfo {volume} {30}},\
  \bibinfo {pages} {1032} (\bibinfo {year} {1984})}\BibitemShut {NoStop}%
\bibitem [{\citenamefont {Ejiri}\ \emph {et~al.}(2019)\citenamefont {Ejiri},
  \citenamefont {Suhonen},\ and\ \citenamefont {Zuber}}]{eji19}%
  \BibitemOpen
  \bibfield  {author} {\bibinfo {author} {\bibfnamefont {H.}~\bibnamefont
  {Ejiri}}, \bibinfo {author} {\bibfnamefont {J.}~\bibnamefont {Suhonen}}, \
  and\ \bibinfo {author} {\bibfnamefont {K.}~\bibnamefont {Zuber}},\ }\href
  {\doibase https://doi.org/10.1016/j.physrep.2018.12.001} {\bibfield
  {journal} {\bibinfo  {journal} {Phys. Rep.}\ }\textbf {\bibinfo {volume}
  {797}},\ \bibinfo {pages} {1 } (\bibinfo {year} {2019})}\BibitemShut
  {NoStop}%
\bibitem [{\citenamefont {Borzov}(2003)}]{bor03}%
  \BibitemOpen
  \bibfield  {author} {\bibinfo {author} {\bibfnamefont {I.~N.}\ \bibnamefont
  {Borzov}},\ }\href {\doibase 10.1103/PhysRevC.67.025802} {\bibfield
  {journal} {\bibinfo  {journal} {Phys. Rev. C}\ }\textbf {\bibinfo {volume}
  {67}},\ \bibinfo {pages} {025802} (\bibinfo {year} {2003})}\BibitemShut
  {NoStop}%
\bibitem [{\citenamefont {Suzuki}\ \emph {et~al.}(2012)\citenamefont {Suzuki},
  \citenamefont {Yoshida}, \citenamefont {Kajino},\ and\ \citenamefont
  {Otsuka}}]{suz12}%
  \BibitemOpen
  \bibfield  {author} {\bibinfo {author} {\bibfnamefont {T.}~\bibnamefont
  {Suzuki}}, \bibinfo {author} {\bibfnamefont {T.}~\bibnamefont {Yoshida}},
  \bibinfo {author} {\bibfnamefont {T.}~\bibnamefont {Kajino}}, \ and\ \bibinfo
  {author} {\bibfnamefont {T.}~\bibnamefont {Otsuka}},\ }\href {\doibase
  10.1103/PhysRevC.85.015802} {\bibfield  {journal} {\bibinfo  {journal} {Phys.
  Rev. C}\ }\textbf {\bibinfo {volume} {85}},\ \bibinfo {pages} {015802}
  (\bibinfo {year} {2012})},\ \Eprint {http://arxiv.org/abs/1110.3886}
  {arXiv:1110.3886 [nucl-th]} \BibitemShut {NoStop}%
\bibitem [{\citenamefont {Zhi}\ \emph {et~al.}(2013)\citenamefont {Zhi},
  \citenamefont {Caurier}, \citenamefont {Cuenca-Garcia}, \citenamefont
  {Langanke}, \citenamefont {Martinez-Pinedo},\ and\ \citenamefont
  {Sieja}}]{zhi13}%
  \BibitemOpen
  \bibfield  {author} {\bibinfo {author} {\bibfnamefont {Q.}~\bibnamefont
  {Zhi}}, \bibinfo {author} {\bibfnamefont {E.}~\bibnamefont {Caurier}},
  \bibinfo {author} {\bibfnamefont {J.}~\bibnamefont {Cuenca-Garcia}}, \bibinfo
  {author} {\bibfnamefont {K.}~\bibnamefont {Langanke}}, \bibinfo {author}
  {\bibfnamefont {G.}~\bibnamefont {Martinez-Pinedo}}, \ and\ \bibinfo {author}
  {\bibfnamefont {K.}~\bibnamefont {Sieja}},\ }\href {\doibase
  10.1103/PhysRevC.87.025803} {\bibfield  {journal} {\bibinfo  {journal} {Phys.
  Rev. C}\ }\textbf {\bibinfo {volume} {87}},\ \bibinfo {pages} {025803}
  (\bibinfo {year} {2013})},\ \Eprint {http://arxiv.org/abs/1301.5225}
  {arXiv:1301.5225 [nucl-th]} \BibitemShut {NoStop}%
\bibitem [{\citenamefont {Fang}\ \emph {et~al.}(2013)\citenamefont {Fang},
  \citenamefont {Brown},\ and\ \citenamefont {Suzuki}}]{fan13}%
  \BibitemOpen
  \bibfield  {author} {\bibinfo {author} {\bibfnamefont {D.-L.}\ \bibnamefont
  {Fang}}, \bibinfo {author} {\bibfnamefont {B.}~\bibnamefont {Brown}}, \ and\
  \bibinfo {author} {\bibfnamefont {T.}~\bibnamefont {Suzuki}},\ }\href
  {\doibase 10.1103/PhysRevC.88.024314} {\bibfield  {journal} {\bibinfo
  {journal} {Phys. Rev. C}\ }\textbf {\bibinfo {volume} {88}},\ \bibinfo
  {pages} {024314} (\bibinfo {year} {2013})},\ \Eprint
  {http://arxiv.org/abs/1211.6070} {arXiv:1211.6070 [nucl-th]} \BibitemShut
  {NoStop}%
\bibitem [{\citenamefont {Mustonen}\ and\ \citenamefont {Engel}(2016)}]{mus16}%
  \BibitemOpen
  \bibfield  {author} {\bibinfo {author} {\bibfnamefont {M.}~\bibnamefont
  {Mustonen}}\ and\ \bibinfo {author} {\bibfnamefont {J.}~\bibnamefont
  {Engel}},\ }\href {\doibase 10.1103/PhysRevC.93.014304} {\bibfield  {journal}
  {\bibinfo  {journal} {Phys. Rev. C}\ }\textbf {\bibinfo {volume} {93}},\
  \bibinfo {pages} {014304} (\bibinfo {year} {2016})},\ \Eprint
  {http://arxiv.org/abs/1510.02136} {arXiv:1510.02136 [nucl-th]} \BibitemShut
  {NoStop}%
\bibitem [{\citenamefont {Marketin}\ \emph {et~al.}(2016)\citenamefont
  {Marketin}, \citenamefont {Huther},\ and\ \citenamefont
  {Mart\'inez-Pinedo}}]{mar16}%
  \BibitemOpen
  \bibfield  {author} {\bibinfo {author} {\bibfnamefont {T.}~\bibnamefont
  {Marketin}}, \bibinfo {author} {\bibfnamefont {L.}~\bibnamefont {Huther}}, \
  and\ \bibinfo {author} {\bibfnamefont {G.}~\bibnamefont
  {Mart\'inez-Pinedo}},\ }\href {\doibase 10.1103/PhysRevC.93.025805}
  {\bibfield  {journal} {\bibinfo  {journal} {Phys. Rev. C}\ }\textbf {\bibinfo
  {volume} {93}},\ \bibinfo {pages} {025805} (\bibinfo {year} {2016})},\
  \Eprint {http://arxiv.org/abs/1507.07442} {arXiv:1507.07442 [nucl-th]}
  \BibitemShut {NoStop}%
\bibitem [{\citenamefont {Yoshida}(2019)}]{yos19}%
  \BibitemOpen
  \bibfield  {author} {\bibinfo {author} {\bibfnamefont {K.}~\bibnamefont
  {Yoshida}},\ }\href {\doibase 10.1103/PhysRevC.100.024316} {\bibfield
  {journal} {\bibinfo  {journal} {Phys. Rev. C}\ }\textbf {\bibinfo {volume}
  {100}},\ \bibinfo {pages} {024316} (\bibinfo {year} {2019})},\ \Eprint
  {http://arxiv.org/abs/1903.03310} {arXiv:1903.03310 [nucl-th]} \BibitemShut
  {NoStop}%
\bibitem [{\citenamefont {Bai}\ \emph {et~al.}(2010)\citenamefont {Bai},
  \citenamefont {Zhang}, \citenamefont {Sagawa}, \citenamefont {Zhang},
  \citenamefont {Col\`o},\ and\ \citenamefont {Xu}}]{bai10}%
  \BibitemOpen
  \bibfield  {author} {\bibinfo {author} {\bibfnamefont {C.}~\bibnamefont
  {Bai}}, \bibinfo {author} {\bibfnamefont {H.}~\bibnamefont {Zhang}}, \bibinfo
  {author} {\bibfnamefont {H.}~\bibnamefont {Sagawa}}, \bibinfo {author}
  {\bibfnamefont {X.}~\bibnamefont {Zhang}}, \bibinfo {author} {\bibfnamefont
  {G.}~\bibnamefont {Col\`o}}, \ and\ \bibinfo {author} {\bibfnamefont
  {F.}~\bibnamefont {Xu}},\ }\href {\doibase 10.1103/PhysRevLett.105.072501}
  {\bibfield  {journal} {\bibinfo  {journal} {Phys. Rev. Lett.}\ }\textbf
  {\bibinfo {volume} {105}},\ \bibinfo {pages} {072501} (\bibinfo {year}
  {2010})},\ \Eprint {http://arxiv.org/abs/1005.2445} {arXiv:1005.2445
  [nucl-th]} \BibitemShut {NoStop}%
\bibitem [{\citenamefont {Horiuchi}\ and\ \citenamefont
  {Suzuki}(2013)}]{hor13}%
  \BibitemOpen
  \bibfield  {author} {\bibinfo {author} {\bibfnamefont {W.}~\bibnamefont
  {Horiuchi}}\ and\ \bibinfo {author} {\bibfnamefont {Y.}~\bibnamefont
  {Suzuki}},\ }\href {\doibase 10.1103/PhysRevC.87.034001} {\bibfield
  {journal} {\bibinfo  {journal} {Phys. Rev. C}\ }\textbf {\bibinfo {volume}
  {87}},\ \bibinfo {pages} {034001} (\bibinfo {year} {2013})},\ \Eprint
  {http://arxiv.org/abs/1301.0196} {arXiv:1301.0196 [nucl-th]} \BibitemShut
  {NoStop}%
\bibitem [{\citenamefont {Yoshida}(2017)}]{yos17}%
  \BibitemOpen
  \bibfield  {author} {\bibinfo {author} {\bibfnamefont {K.}~\bibnamefont
  {Yoshida}},\ }\href {\doibase 10.1103/PhysRevC.96.051302} {\bibfield
  {journal} {\bibinfo  {journal} {Phys. Rev. C}\ }\textbf {\bibinfo {volume}
  {96}},\ \bibinfo {pages} {051302} (\bibinfo {year} {2017})},\ \Eprint
  {http://arxiv.org/abs/1709.10272} {arXiv:1709.10272 [nucl-th]} \BibitemShut
  {NoStop}%
\bibitem [{\citenamefont {Bender}\ \emph {et~al.}(2003)\citenamefont {Bender},
  \citenamefont {Heenen},\ and\ \citenamefont {Reinhard}}]{ben03}%
  \BibitemOpen
  \bibfield  {author} {\bibinfo {author} {\bibfnamefont {M.}~\bibnamefont
  {Bender}}, \bibinfo {author} {\bibfnamefont {P.-H.}\ \bibnamefont {Heenen}},
  \ and\ \bibinfo {author} {\bibfnamefont {P.-G.}\ \bibnamefont {Reinhard}},\
  }\href {\doibase 10.1103/RevModPhys.75.121} {\bibfield  {journal} {\bibinfo
  {journal} {Rev. Mod. Phys.}\ }\textbf {\bibinfo {volume} {75}},\ \bibinfo
  {pages} {121} (\bibinfo {year} {2003})}\BibitemShut {NoStop}%
\bibitem [{\citenamefont {Nakatsukasa}\ \emph {et~al.}(2016)\citenamefont
  {Nakatsukasa}, \citenamefont {Matsuyanagi}, \citenamefont {Matsuo},\ and\
  \citenamefont {Yabana}}]{nak16}%
  \BibitemOpen
  \bibfield  {author} {\bibinfo {author} {\bibfnamefont {T.}~\bibnamefont
  {Nakatsukasa}}, \bibinfo {author} {\bibfnamefont {K.}~\bibnamefont
  {Matsuyanagi}}, \bibinfo {author} {\bibfnamefont {M.}~\bibnamefont {Matsuo}},
  \ and\ \bibinfo {author} {\bibfnamefont {K.}~\bibnamefont {Yabana}},\ }\href
  {\doibase 10.1103/RevModPhys.88.045004} {\bibfield  {journal} {\bibinfo
  {journal} {Rev. Mod. Phys.}\ }\textbf {\bibinfo {volume} {88}},\ \bibinfo
  {pages} {045004} (\bibinfo {year} {2016})},\ \Eprint
  {http://arxiv.org/abs/1606.04717} {arXiv:1606.04717} \BibitemShut {NoStop}%
\bibitem [{\citenamefont {Yoshida}(2013)}]{yos13}%
  \BibitemOpen
  \bibfield  {author} {\bibinfo {author} {\bibfnamefont {K.}~\bibnamefont
  {Yoshida}},\ }\href {\doibase 10.1093/ptep/ptt091} {\bibfield  {journal}
  {\bibinfo  {journal} {Prog. Theor. Exp. Phys.}\ }\textbf {\bibinfo {volume}
  {2013}},\ \bibinfo {pages} {113D02} (\bibinfo {year} {2013})},\ \Eprint
  {http://arxiv.org/abs/1308.0424} {arXiv:1308.0424 [nucl-th]} \BibitemShut
  {NoStop}%
\bibitem [{\citenamefont {Dobaczewski}\ \emph {et~al.}(1984)\citenamefont
  {Dobaczewski}, \citenamefont {Flocard},\ and\ \citenamefont
  {Treiner}}]{dob84}%
  \BibitemOpen
  \bibfield  {author} {\bibinfo {author} {\bibfnamefont {J.}~\bibnamefont
  {Dobaczewski}}, \bibinfo {author} {\bibfnamefont {H.}~\bibnamefont
  {Flocard}}, \ and\ \bibinfo {author} {\bibfnamefont {J.}~\bibnamefont
  {Treiner}},\ }\href {\doibase https://doi.org/10.1016/0375-9474(84)90433-0}
  {\bibfield  {journal} {\bibinfo  {journal} {Nucl. Phys. A}\ }\textbf
  {\bibinfo {volume} {422}},\ \bibinfo {pages} {103 } (\bibinfo {year}
  {1984})}\BibitemShut {NoStop}%
\bibitem [{\citenamefont {Bartel}\ \emph {et~al.}(1982)\citenamefont {Bartel},
  \citenamefont {Quentin}, \citenamefont {Brack}, \citenamefont {Guet},\ and\
  \citenamefont {H\r{a}kansson}}]{bar82}%
  \BibitemOpen
  \bibfield  {author} {\bibinfo {author} {\bibfnamefont {J.}~\bibnamefont
  {Bartel}}, \bibinfo {author} {\bibfnamefont {P.}~\bibnamefont {Quentin}},
  \bibinfo {author} {\bibfnamefont {M.}~\bibnamefont {Brack}}, \bibinfo
  {author} {\bibfnamefont {C.}~\bibnamefont {Guet}}, \ and\ \bibinfo {author}
  {\bibfnamefont {H.-B.}\ \bibnamefont {H\r{a}kansson}},\ }\href {\doibase
  https://doi.org/10.1016/0375-9474(82)90403-1} {\bibfield  {journal} {\bibinfo
   {journal} {Nucl. Phys. A}\ }\textbf {\bibinfo {volume} {386}},\ \bibinfo
  {pages} {79 } (\bibinfo {year} {1982})}\BibitemShut {NoStop}%
\bibitem [{\citenamefont {Yamagami}\ \emph {et~al.}(2009)\citenamefont
  {Yamagami}, \citenamefont {Shimizu},\ and\ \citenamefont
  {Nakatsukasa}}]{yam09}%
  \BibitemOpen
  \bibfield  {author} {\bibinfo {author} {\bibfnamefont {M.}~\bibnamefont
  {Yamagami}}, \bibinfo {author} {\bibfnamefont {Y.}~\bibnamefont {Shimizu}}, \
  and\ \bibinfo {author} {\bibfnamefont {T.}~\bibnamefont {Nakatsukasa}},\
  }\href {\doibase 10.1103/PhysRevC.80.064301} {\bibfield  {journal} {\bibinfo
  {journal} {Phys. Rev. C}\ }\textbf {\bibinfo {volume} {80}},\ \bibinfo
  {pages} {064301} (\bibinfo {year} {2009})},\ \Eprint
  {http://arxiv.org/abs/0812.3197} {arXiv:0812.3197 [nucl-th]} \BibitemShut
  {NoStop}%
\bibitem [{\citenamefont {Frauendorf}\ and\ \citenamefont
  {Macchiavelli}(2014)}]{fra14}%
  \BibitemOpen
  \bibfield  {author} {\bibinfo {author} {\bibfnamefont {S.}~\bibnamefont
  {Frauendorf}}\ and\ \bibinfo {author} {\bibfnamefont {A.~O.}\ \bibnamefont
  {Macchiavelli}},\ }\href {\doibase 10.1016/j.ppnp.2014.07.001} {\bibfield
  {journal} {\bibinfo  {journal} {Prog. Part. Nucl. Phys.}\ }\textbf {\bibinfo
  {volume} {78}},\ \bibinfo {pages} {24} (\bibinfo {year} {2014})},\ \Eprint
  {http://arxiv.org/abs/1405.1652} {arXiv:1405.1652 [nucl-th]} \BibitemShut
  {NoStop}%
\bibitem [{\citenamefont {Fujita}\ \emph {et~al.}(2014)\citenamefont {Fujita},
  \citenamefont {Fujita}, \citenamefont {Adachi}, \citenamefont {Bai},
  \citenamefont {Algora}, \citenamefont {Berg}, \citenamefont {von Brentano},
  \citenamefont {Col\`o}, \citenamefont {Csatl\'os}, \citenamefont {Deaven},
  \citenamefont {Estevez-Aguado}, \citenamefont {Fransen}, \citenamefont
  {De~Frenne}, \citenamefont {Fujita}, \citenamefont
  {Ganio\ifmmode~\breve{g}\else \u{g}\fi{}lu}, \citenamefont {Guess},
  \citenamefont {Guly\'as}, \citenamefont {Hatanaka}, \citenamefont {Hirota},
  \citenamefont {Honma}, \citenamefont {Ishikawa}, \citenamefont {Jacobs},
  \citenamefont {Krasznahorkay}, \citenamefont {Matsubara}, \citenamefont
  {Matsuyanagi}, \citenamefont {Meharchand}, \citenamefont {Molina},
  \citenamefont {Muto}, \citenamefont {Nakanishi}, \citenamefont {Negret},
  \citenamefont {Okamura}, \citenamefont {Ong}, \citenamefont {Otsuka},
  \citenamefont {Pietralla}, \citenamefont {Perdikakis}, \citenamefont
  {Popescu}, \citenamefont {Rubio}, \citenamefont {Sagawa}, \citenamefont
  {Sarriguren}, \citenamefont {Scholl}, \citenamefont {Shimbara}, \citenamefont
  {Shimizu}, \citenamefont {Susoy}, \citenamefont {Suzuki}, \citenamefont
  {Tameshige}, \citenamefont {Tamii}, \citenamefont {Thies}, \citenamefont
  {Uchida}, \citenamefont {Wakasa}, \citenamefont {Yosoi}, \citenamefont
  {Zegers}, \citenamefont {Zell},\ and\ \citenamefont {Zenihiro}}]{fuj14}%
  \BibitemOpen
  \bibfield  {author} {\bibinfo {author} {\bibfnamefont {Y.}~\bibnamefont
  {Fujita}}, \bibinfo {author} {\bibfnamefont {H.}~\bibnamefont {Fujita}},
  \bibinfo {author} {\bibfnamefont {T.}~\bibnamefont {Adachi}}, \bibinfo
  {author} {\bibfnamefont {C.~L.}\ \bibnamefont {Bai}}, \bibinfo {author}
  {\bibfnamefont {A.}~\bibnamefont {Algora}}, \bibinfo {author} {\bibfnamefont
  {G.~P.~A.}\ \bibnamefont {Berg}}, \bibinfo {author} {\bibfnamefont
  {P.}~\bibnamefont {von Brentano}}, \bibinfo {author} {\bibfnamefont
  {G.}~\bibnamefont {Col\`o}}, \bibinfo {author} {\bibfnamefont
  {M.}~\bibnamefont {Csatl\'os}}, \bibinfo {author} {\bibfnamefont {J.~M.}\
  \bibnamefont {Deaven}}, \bibinfo {author} {\bibfnamefont {E.}~\bibnamefont
  {Estevez-Aguado}}, \bibinfo {author} {\bibfnamefont {C.}~\bibnamefont
  {Fransen}}, \bibinfo {author} {\bibfnamefont {D.}~\bibnamefont {De~Frenne}},
  \bibinfo {author} {\bibfnamefont {K.}~\bibnamefont {Fujita}}, \bibinfo
  {author} {\bibfnamefont {E.}~\bibnamefont {Ganio\ifmmode~\breve{g}\else
  \u{g}\fi{}lu}}, \bibinfo {author} {\bibfnamefont {C.~J.}\ \bibnamefont
  {Guess}}, \bibinfo {author} {\bibfnamefont {J.}~\bibnamefont {Guly\'as}},
  \bibinfo {author} {\bibfnamefont {K.}~\bibnamefont {Hatanaka}}, \bibinfo
  {author} {\bibfnamefont {K.}~\bibnamefont {Hirota}}, \bibinfo {author}
  {\bibfnamefont {M.}~\bibnamefont {Honma}}, \bibinfo {author} {\bibfnamefont
  {D.}~\bibnamefont {Ishikawa}}, \bibinfo {author} {\bibfnamefont
  {E.}~\bibnamefont {Jacobs}}, \bibinfo {author} {\bibfnamefont
  {A.}~\bibnamefont {Krasznahorkay}}, \bibinfo {author} {\bibfnamefont
  {H.}~\bibnamefont {Matsubara}}, \bibinfo {author} {\bibfnamefont
  {K.}~\bibnamefont {Matsuyanagi}}, \bibinfo {author} {\bibfnamefont
  {R.}~\bibnamefont {Meharchand}}, \bibinfo {author} {\bibfnamefont
  {F.}~\bibnamefont {Molina}}, \bibinfo {author} {\bibfnamefont
  {K.}~\bibnamefont {Muto}}, \bibinfo {author} {\bibfnamefont {K.}~\bibnamefont
  {Nakanishi}}, \bibinfo {author} {\bibfnamefont {A.}~\bibnamefont {Negret}},
  \bibinfo {author} {\bibfnamefont {H.}~\bibnamefont {Okamura}}, \bibinfo
  {author} {\bibfnamefont {H.~J.}\ \bibnamefont {Ong}}, \bibinfo {author}
  {\bibfnamefont {T.}~\bibnamefont {Otsuka}}, \bibinfo {author} {\bibfnamefont
  {N.}~\bibnamefont {Pietralla}}, \bibinfo {author} {\bibfnamefont
  {G.}~\bibnamefont {Perdikakis}}, \bibinfo {author} {\bibfnamefont
  {L.}~\bibnamefont {Popescu}}, \bibinfo {author} {\bibfnamefont
  {B.}~\bibnamefont {Rubio}}, \bibinfo {author} {\bibfnamefont
  {H.}~\bibnamefont {Sagawa}}, \bibinfo {author} {\bibfnamefont
  {P.}~\bibnamefont {Sarriguren}}, \bibinfo {author} {\bibfnamefont
  {C.}~\bibnamefont {Scholl}}, \bibinfo {author} {\bibfnamefont
  {Y.}~\bibnamefont {Shimbara}}, \bibinfo {author} {\bibfnamefont
  {Y.}~\bibnamefont {Shimizu}}, \bibinfo {author} {\bibfnamefont
  {G.}~\bibnamefont {Susoy}}, \bibinfo {author} {\bibfnamefont
  {T.}~\bibnamefont {Suzuki}}, \bibinfo {author} {\bibfnamefont
  {Y.}~\bibnamefont {Tameshige}}, \bibinfo {author} {\bibfnamefont
  {A.}~\bibnamefont {Tamii}}, \bibinfo {author} {\bibfnamefont {J.~H.}\
  \bibnamefont {Thies}}, \bibinfo {author} {\bibfnamefont {M.}~\bibnamefont
  {Uchida}}, \bibinfo {author} {\bibfnamefont {T.}~\bibnamefont {Wakasa}},
  \bibinfo {author} {\bibfnamefont {M.}~\bibnamefont {Yosoi}}, \bibinfo
  {author} {\bibfnamefont {R.~G.~T.}\ \bibnamefont {Zegers}}, \bibinfo {author}
  {\bibfnamefont {K.~O.}\ \bibnamefont {Zell}}, \ and\ \bibinfo {author}
  {\bibfnamefont {J.}~\bibnamefont {Zenihiro}},\ }\href {\doibase
  10.1103/PhysRevLett.112.112502} {\bibfield  {journal} {\bibinfo  {journal}
  {Phys. Rev. Lett.}\ }\textbf {\bibinfo {volume} {112}},\ \bibinfo {pages}
  {112502} (\bibinfo {year} {2014})}\BibitemShut {NoStop}%
\bibitem [{\citenamefont {Fujita}\ \emph {et~al.}(2015)\citenamefont {Fujita},
  \citenamefont {Fujita}, \citenamefont {Adachi}, \citenamefont {Susoy},
  \citenamefont {Algora}, \citenamefont {Bai}, \citenamefont {Col\`o},
  \citenamefont {Csatl\'os}, \citenamefont {Deaven}, \citenamefont
  {Estevez-Aguado}, \citenamefont {Guess}, \citenamefont {Guly\'as},
  \citenamefont {Hatanaka}, \citenamefont {Hirota}, \citenamefont {Honma},
  \citenamefont {Ishikawa}, \citenamefont {Krasznahorkay}, \citenamefont
  {Matsubara}, \citenamefont {Meharchand}, \citenamefont {Molina},
  \citenamefont {Nakada}, \citenamefont {Okamura}, \citenamefont {Ong},
  \citenamefont {Otsuka}, \citenamefont {Perdikakis}, \citenamefont {Rubio},
  \citenamefont {Sagawa}, \citenamefont {Sarriguren}, \citenamefont {Scholl},
  \citenamefont {Shimbara}, \citenamefont {Stephenson}, \citenamefont {Suzuki},
  \citenamefont {Tamii}, \citenamefont {Thies}, \citenamefont {Yoshida},
  \citenamefont {Zegers},\ and\ \citenamefont {Zenihiro}}]{fuj15}%
  \BibitemOpen
  \bibfield  {author} {\bibinfo {author} {\bibfnamefont {Y.}~\bibnamefont
  {Fujita}}, \bibinfo {author} {\bibfnamefont {H.}~\bibnamefont {Fujita}},
  \bibinfo {author} {\bibfnamefont {T.}~\bibnamefont {Adachi}}, \bibinfo
  {author} {\bibfnamefont {G.}~\bibnamefont {Susoy}}, \bibinfo {author}
  {\bibfnamefont {A.}~\bibnamefont {Algora}}, \bibinfo {author} {\bibfnamefont
  {C.~L.}\ \bibnamefont {Bai}}, \bibinfo {author} {\bibfnamefont
  {G.}~\bibnamefont {Col\`o}}, \bibinfo {author} {\bibfnamefont
  {M.}~\bibnamefont {Csatl\'os}}, \bibinfo {author} {\bibfnamefont {J.~M.}\
  \bibnamefont {Deaven}}, \bibinfo {author} {\bibfnamefont {E.}~\bibnamefont
  {Estevez-Aguado}}, \bibinfo {author} {\bibfnamefont {C.~J.}\ \bibnamefont
  {Guess}}, \bibinfo {author} {\bibfnamefont {J.}~\bibnamefont {Guly\'as}},
  \bibinfo {author} {\bibfnamefont {K.}~\bibnamefont {Hatanaka}}, \bibinfo
  {author} {\bibfnamefont {K.}~\bibnamefont {Hirota}}, \bibinfo {author}
  {\bibfnamefont {M.}~\bibnamefont {Honma}}, \bibinfo {author} {\bibfnamefont
  {D.}~\bibnamefont {Ishikawa}}, \bibinfo {author} {\bibfnamefont
  {A.}~\bibnamefont {Krasznahorkay}}, \bibinfo {author} {\bibfnamefont
  {H.}~\bibnamefont {Matsubara}}, \bibinfo {author} {\bibfnamefont
  {R.}~\bibnamefont {Meharchand}}, \bibinfo {author} {\bibfnamefont
  {F.}~\bibnamefont {Molina}}, \bibinfo {author} {\bibfnamefont
  {H.}~\bibnamefont {Nakada}}, \bibinfo {author} {\bibfnamefont
  {H.}~\bibnamefont {Okamura}}, \bibinfo {author} {\bibfnamefont {H.~J.}\
  \bibnamefont {Ong}}, \bibinfo {author} {\bibfnamefont {T.}~\bibnamefont
  {Otsuka}}, \bibinfo {author} {\bibfnamefont {G.}~\bibnamefont {Perdikakis}},
  \bibinfo {author} {\bibfnamefont {B.}~\bibnamefont {Rubio}}, \bibinfo
  {author} {\bibfnamefont {H.}~\bibnamefont {Sagawa}}, \bibinfo {author}
  {\bibfnamefont {P.}~\bibnamefont {Sarriguren}}, \bibinfo {author}
  {\bibfnamefont {C.}~\bibnamefont {Scholl}}, \bibinfo {author} {\bibfnamefont
  {Y.}~\bibnamefont {Shimbara}}, \bibinfo {author} {\bibfnamefont {E.~J.}\
  \bibnamefont {Stephenson}}, \bibinfo {author} {\bibfnamefont
  {T.}~\bibnamefont {Suzuki}}, \bibinfo {author} {\bibfnamefont
  {A.}~\bibnamefont {Tamii}}, \bibinfo {author} {\bibfnamefont {J.~H.}\
  \bibnamefont {Thies}}, \bibinfo {author} {\bibfnamefont {K.}~\bibnamefont
  {Yoshida}}, \bibinfo {author} {\bibfnamefont {R.~G.~T.}\ \bibnamefont
  {Zegers}}, \ and\ \bibinfo {author} {\bibfnamefont {J.}~\bibnamefont
  {Zenihiro}},\ }\href {\doibase 10.1103/PhysRevC.91.064316} {\bibfield
  {journal} {\bibinfo  {journal} {Phys. Rev. C}\ }\textbf {\bibinfo {volume}
  {91}},\ \bibinfo {pages} {064316} (\bibinfo {year} {2015})}\BibitemShut
  {NoStop}%
\bibitem [{\citenamefont {Fujita}\ \emph {et~al.}(2019)\citenamefont {Fujita},
  \citenamefont {Fujita}, \citenamefont {Utsuno}, \citenamefont {Yoshida},
  \citenamefont {Adachi}, \citenamefont {Algora}, \citenamefont {Csatl\'os},
  \citenamefont {Deaven}, \citenamefont {Estevez-Aguado}, \citenamefont
  {Guess}, \citenamefont {Guly\'as}, \citenamefont {Hatanaka}, \citenamefont
  {Hirota}, \citenamefont {Hutton}, \citenamefont {Ishikawa}, \citenamefont
  {Krasznahorkay}, \citenamefont {Matsubara}, \citenamefont {Molina},
  \citenamefont {Okamura}, \citenamefont {Ong}, \citenamefont {Perdikakis},
  \citenamefont {Rubio}, \citenamefont {Scholl}, \citenamefont {Shimbara},
  \citenamefont {S\"usoy}, \citenamefont {Suzuki}, \citenamefont {Tamii},
  \citenamefont {Thies}, \citenamefont {Zegers},\ and\ \citenamefont
  {Zenihiro}}]{fuj19}%
  \BibitemOpen
  \bibfield  {author} {\bibinfo {author} {\bibfnamefont {H.}~\bibnamefont
  {Fujita}}, \bibinfo {author} {\bibfnamefont {Y.}~\bibnamefont {Fujita}},
  \bibinfo {author} {\bibfnamefont {Y.}~\bibnamefont {Utsuno}}, \bibinfo
  {author} {\bibfnamefont {K.}~\bibnamefont {Yoshida}}, \bibinfo {author}
  {\bibfnamefont {T.}~\bibnamefont {Adachi}}, \bibinfo {author} {\bibfnamefont
  {A.}~\bibnamefont {Algora}}, \bibinfo {author} {\bibfnamefont
  {M.}~\bibnamefont {Csatl\'os}}, \bibinfo {author} {\bibfnamefont {J.~M.}\
  \bibnamefont {Deaven}}, \bibinfo {author} {\bibfnamefont {E.}~\bibnamefont
  {Estevez-Aguado}}, \bibinfo {author} {\bibfnamefont {C.~J.}\ \bibnamefont
  {Guess}}, \bibinfo {author} {\bibfnamefont {J.}~\bibnamefont {Guly\'as}},
  \bibinfo {author} {\bibfnamefont {K.}~\bibnamefont {Hatanaka}}, \bibinfo
  {author} {\bibfnamefont {K.}~\bibnamefont {Hirota}}, \bibinfo {author}
  {\bibfnamefont {R.}~\bibnamefont {Hutton}}, \bibinfo {author} {\bibfnamefont
  {D.}~\bibnamefont {Ishikawa}}, \bibinfo {author} {\bibfnamefont
  {A.}~\bibnamefont {Krasznahorkay}}, \bibinfo {author} {\bibfnamefont
  {H.}~\bibnamefont {Matsubara}}, \bibinfo {author} {\bibfnamefont
  {F.}~\bibnamefont {Molina}}, \bibinfo {author} {\bibfnamefont
  {H.}~\bibnamefont {Okamura}}, \bibinfo {author} {\bibfnamefont {H.~J.}\
  \bibnamefont {Ong}}, \bibinfo {author} {\bibfnamefont {G.}~\bibnamefont
  {Perdikakis}}, \bibinfo {author} {\bibfnamefont {B.}~\bibnamefont {Rubio}},
  \bibinfo {author} {\bibfnamefont {C.}~\bibnamefont {Scholl}}, \bibinfo
  {author} {\bibfnamefont {Y.}~\bibnamefont {Shimbara}}, \bibinfo {author}
  {\bibfnamefont {G.}~\bibnamefont {S\"usoy}}, \bibinfo {author} {\bibfnamefont
  {T.}~\bibnamefont {Suzuki}}, \bibinfo {author} {\bibfnamefont
  {A.}~\bibnamefont {Tamii}}, \bibinfo {author} {\bibfnamefont {J.~H.}\
  \bibnamefont {Thies}}, \bibinfo {author} {\bibfnamefont {R.~G.~T.}\
  \bibnamefont {Zegers}}, \ and\ \bibinfo {author} {\bibfnamefont
  {J.}~\bibnamefont {Zenihiro}},\ }\href {\doibase 10.1103/PhysRevC.100.034618}
  {\bibfield  {journal} {\bibinfo  {journal} {Phys. Rev. C}\ }\textbf {\bibinfo
  {volume} {100}},\ \bibinfo {pages} {034618} (\bibinfo {year}
  {2019})}\BibitemShut {NoStop}%
\bibitem [{\citenamefont {Yoshida}\ and\ \citenamefont
  {Van~Giai}(2008)}]{yos08}%
  \BibitemOpen
  \bibfield  {author} {\bibinfo {author} {\bibfnamefont {K.}~\bibnamefont
  {Yoshida}}\ and\ \bibinfo {author} {\bibfnamefont {N.}~\bibnamefont
  {Van~Giai}},\ }\href {\doibase 10.1103/PhysRevC.78.064316} {\bibfield
  {journal} {\bibinfo  {journal} {Phys. Rev. C}\ }\textbf {\bibinfo {volume}
  {78}},\ \bibinfo {pages} {064316} (\bibinfo {year} {2008})},\ \Eprint
  {http://arxiv.org/abs/0809.0169} {arXiv:0809.0169 [nucl-th]} \BibitemShut
  {NoStop}%
\bibitem [{\citenamefont {Yoshida}\ and\ \citenamefont
  {Nakatsukasa}(2013)}]{yos13b}%
  \BibitemOpen
  \bibfield  {author} {\bibinfo {author} {\bibfnamefont {K.}~\bibnamefont
  {Yoshida}}\ and\ \bibinfo {author} {\bibfnamefont {T.}~\bibnamefont
  {Nakatsukasa}},\ }\href {\doibase 10.1103/PhysRevC.88.034309} {\bibfield
  {journal} {\bibinfo  {journal} {Phys. Rev. C}\ }\textbf {\bibinfo {volume}
  {88}},\ \bibinfo {pages} {034309} (\bibinfo {year} {2013})},\ \Eprint
  {http://arxiv.org/abs/1305.6437} {arXiv:1305.6437 [nucl-th]} \BibitemShut
  {NoStop}%
\bibitem [{\citenamefont {Ring}\ and\ \citenamefont {Schuck}(1980)}]{rin80}%
  \BibitemOpen
  \bibfield  {author} {\bibinfo {author} {\bibfnamefont {P.}~\bibnamefont
  {Ring}}\ and\ \bibinfo {author} {\bibfnamefont {P.}~\bibnamefont {Schuck}},\
  }\href@noop {} {\emph {\bibinfo {title} {The nuclear many-body problem}}}\
  (\bibinfo  {publisher} {Springer-Verlag},\ \bibinfo {address} {New York},\
  \bibinfo {year} {1980})\BibitemShut {NoStop}%
\bibitem [{\citenamefont {Kleinig}\ \emph {et~al.}(2008)\citenamefont
  {Kleinig}, \citenamefont {Nesterenko}, \citenamefont {Kvasil}, \citenamefont
  {Reinhard},\ and\ \citenamefont {Vesely}}]{kle08}%
  \BibitemOpen
  \bibfield  {author} {\bibinfo {author} {\bibfnamefont {W.}~\bibnamefont
  {Kleinig}}, \bibinfo {author} {\bibfnamefont {V.}~\bibnamefont {Nesterenko}},
  \bibinfo {author} {\bibfnamefont {J.}~\bibnamefont {Kvasil}}, \bibinfo
  {author} {\bibfnamefont {P.-G.}\ \bibnamefont {Reinhard}}, \ and\ \bibinfo
  {author} {\bibfnamefont {P.}~\bibnamefont {Vesely}},\ }\href {\doibase
  10.1103/PhysRevC.78.044313} {\bibfield  {journal} {\bibinfo  {journal} {Phys.
  Rev. C}\ }\textbf {\bibinfo {volume} {78}},\ \bibinfo {pages} {044313}
  (\bibinfo {year} {2008})},\ \Eprint {http://arxiv.org/abs/0805.4787}
  {arXiv:0805.4787 [nucl-th]} \BibitemShut {NoStop}%
\bibitem [{\citenamefont {Yoshida}\ and\ \citenamefont
  {Nakatsukasa}(2011)}]{yos11b}%
  \BibitemOpen
  \bibfield  {author} {\bibinfo {author} {\bibfnamefont {K.}~\bibnamefont
  {Yoshida}}\ and\ \bibinfo {author} {\bibfnamefont {T.}~\bibnamefont
  {Nakatsukasa}},\ }\href {\doibase 10.1103/PhysRevC.83.021304} {\bibfield
  {journal} {\bibinfo  {journal} {Phys. Rev. C}\ }\textbf {\bibinfo {volume}
  {83}},\ \bibinfo {pages} {021304} (\bibinfo {year} {2011})},\ \Eprint
  {http://arxiv.org/abs/1008.1520} {arXiv:1008.1520 [nucl-th]} \BibitemShut
  {NoStop}%
\bibitem [{\citenamefont {Stetcu}\ \emph {et~al.}(2011)\citenamefont {Stetcu},
  \citenamefont {Bulgac}, \citenamefont {Magierski},\ and\ \citenamefont
  {Roche}}]{ste11}%
  \BibitemOpen
  \bibfield  {author} {\bibinfo {author} {\bibfnamefont {I.}~\bibnamefont
  {Stetcu}}, \bibinfo {author} {\bibfnamefont {A.}~\bibnamefont {Bulgac}},
  \bibinfo {author} {\bibfnamefont {P.}~\bibnamefont {Magierski}}, \ and\
  \bibinfo {author} {\bibfnamefont {K.}~\bibnamefont {Roche}},\ }\href
  {\doibase 10.1103/PhysRevC.84.051309} {\bibfield  {journal} {\bibinfo
  {journal} {Phys. Rev. C}\ }\textbf {\bibinfo {volume} {84}},\ \bibinfo
  {pages} {051309} (\bibinfo {year} {2011})},\ \Eprint
  {http://arxiv.org/abs/1108.3064} {arXiv:1108.3064 [nucl-th]} \BibitemShut
  {NoStop}%
\bibitem [{\citenamefont {Nakatsukasa}\ \emph {et~al.}(2011)\citenamefont
  {Nakatsukasa}, \citenamefont {Avogadro}, \citenamefont {Ebata}, \citenamefont
  {Inakura},\ and\ \citenamefont {Yoshida}}]{nak11}%
  \BibitemOpen
  \bibfield  {author} {\bibinfo {author} {\bibfnamefont {T.}~\bibnamefont
  {Nakatsukasa}}, \bibinfo {author} {\bibfnamefont {P.}~\bibnamefont
  {Avogadro}}, \bibinfo {author} {\bibfnamefont {S.}~\bibnamefont {Ebata}},
  \bibinfo {author} {\bibfnamefont {T.}~\bibnamefont {Inakura}}, \ and\
  \bibinfo {author} {\bibfnamefont {K.}~\bibnamefont {Yoshida}},\ }\href
  {\doibase 10.5506/APhysPolB.42.609} {\bibfield  {journal} {\bibinfo
  {journal} {Acta Phys. Polon. B}\ }\textbf {\bibinfo {volume} {42}},\ \bibinfo
  {pages} {609} (\bibinfo {year} {2011})},\ \Eprint
  {http://arxiv.org/abs/1101.3106} {arXiv:1101.3106 [nucl-th]} \BibitemShut
  {NoStop}%
\bibitem [{\citenamefont {Oishi}\ \emph {et~al.}(2016)\citenamefont {Oishi},
  \citenamefont {Kortelainen},\ and\ \citenamefont {Hinohara}}]{ois16}%
  \BibitemOpen
  \bibfield  {author} {\bibinfo {author} {\bibfnamefont {T.}~\bibnamefont
  {Oishi}}, \bibinfo {author} {\bibfnamefont {M.}~\bibnamefont {Kortelainen}},
  \ and\ \bibinfo {author} {\bibfnamefont {N.}~\bibnamefont {Hinohara}},\
  }\href {\doibase 10.1103/PhysRevC.93.034329} {\bibfield  {journal} {\bibinfo
  {journal} {Phys. Rev. C}\ }\textbf {\bibinfo {volume} {93}},\ \bibinfo
  {pages} {034329} (\bibinfo {year} {2016})},\ \Eprint
  {http://arxiv.org/abs/1512.09146} {arXiv:1512.09146 [nucl-th]} \BibitemShut
  {NoStop}%
\bibitem [{\citenamefont {Mennana}\ \emph {et~al.}(2020)\citenamefont
  {Mennana}, \citenamefont {Bassem},\ and\ \citenamefont {Oulne}}]{men20}%
  \BibitemOpen
  \bibfield  {author} {\bibinfo {author} {\bibfnamefont {A.~A.~B.}\
  \bibnamefont {Mennana}}, \bibinfo {author} {\bibfnamefont {Y.}~\bibnamefont
  {Bassem}}, \ and\ \bibinfo {author} {\bibfnamefont {M.}~\bibnamefont
  {Oulne}},\ }\href {\doibase 10.1088/1402-4896/ab73d8} {\bibfield  {journal}
  {\bibinfo  {journal} {Phys. Scr.}\ }\textbf {\bibinfo {volume} {95}},\
  \bibinfo {pages} {065301} (\bibinfo {year} {2020})},\ \Eprint
  {http://arxiv.org/abs/2005.04337} {arXiv:2005.04337 [nucl-th]} \BibitemShut
  {NoStop}%
\bibitem [{\citenamefont {Chabanat}\ \emph {et~al.}(1998)\citenamefont
  {Chabanat}, \citenamefont {Bonche}, \citenamefont {Haensel}, \citenamefont
  {Meyer},\ and\ \citenamefont {Schaeffer}}]{cha98}%
  \BibitemOpen
  \bibfield  {author} {\bibinfo {author} {\bibfnamefont {E.}~\bibnamefont
  {Chabanat}}, \bibinfo {author} {\bibfnamefont {P.}~\bibnamefont {Bonche}},
  \bibinfo {author} {\bibfnamefont {P.}~\bibnamefont {Haensel}}, \bibinfo
  {author} {\bibfnamefont {J.}~\bibnamefont {Meyer}}, \ and\ \bibinfo {author}
  {\bibfnamefont {R.}~\bibnamefont {Schaeffer}},\ }\href {\doibase
  10.1016/S0375-9474(98)00180-8} {\bibfield  {journal} {\bibinfo  {journal}
  {Nucl. Phys. A}\ }\textbf {\bibinfo {volume} {635}},\ \bibinfo {pages} {231}
  (\bibinfo {year} {1998})},\ \bibinfo {note} {[Erratum: Nucl.Phys.A 643,
  441--441 (1998)]}\BibitemShut {NoStop}%
\bibitem [{\citenamefont {Bertsch}\ and\ \citenamefont
  {Broglia}(1994)}]{ber94}%
  \BibitemOpen
  \bibfield  {author} {\bibinfo {author} {\bibfnamefont {G.~F.}\ \bibnamefont
  {Bertsch}}\ and\ \bibinfo {author} {\bibfnamefont {R.~A.}\ \bibnamefont
  {Broglia}},\ }\href {http://www.loc.gov/catdir/toc/cam028/92040596.html
  http://www.loc.gov/catdir/samples/cam031/92040596.html} {\emph {\bibinfo
  {title} {Oscillations in finite quantum systems}}}\ (\bibinfo  {publisher}
  {Cambridge University Press Cambridge [England] ; New York},\ \bibinfo {year}
  {1994})\BibitemShut {NoStop}%
\bibitem [{\citenamefont {van Giai}\ and\ \citenamefont
  {Sagawa}(1981)}]{gia81}%
  \BibitemOpen
  \bibfield  {author} {\bibinfo {author} {\bibfnamefont {N.}~\bibnamefont {van
  Giai}}\ and\ \bibinfo {author} {\bibfnamefont {H.}~\bibnamefont {Sagawa}},\
  }\href {\doibase 10.1016/0370-2693(81)90646-8} {\bibfield  {journal}
  {\bibinfo  {journal} {Phys. Lett. B}\ }\textbf {\bibinfo {volume} {106}},\
  \bibinfo {pages} {379} (\bibinfo {year} {1981})}\BibitemShut {NoStop}%
\bibitem [{\citenamefont {Gurevich}\ \emph {et~al.}(1976)\citenamefont
  {Gurevich}, \citenamefont {Lazareva}, \citenamefont {Mazur}, \citenamefont
  {Solodukhov},\ and\ \citenamefont {Tulupov}}]{gur76}%
  \BibitemOpen
  \bibfield  {author} {\bibinfo {author} {\bibfnamefont {G.}~\bibnamefont
  {Gurevich}}, \bibinfo {author} {\bibfnamefont {L.}~\bibnamefont {Lazareva}},
  \bibinfo {author} {\bibfnamefont {V.}~\bibnamefont {Mazur}}, \bibinfo
  {author} {\bibfnamefont {G.}~\bibnamefont {Solodukhov}}, \ and\ \bibinfo
  {author} {\bibfnamefont {B.}~\bibnamefont {Tulupov}},\ }\href {\doibase
  https://doi.org/10.1016/0375-9474(76)90594-7} {\bibfield  {journal} {\bibinfo
   {journal} {Nucl. Phys. A}\ }\textbf {\bibinfo {volume} {273}},\ \bibinfo
  {pages} {326 } (\bibinfo {year} {1976})}\BibitemShut {NoStop}%
\bibitem [{\citenamefont {Yoshida}\ \emph {et~al.}(2011)\citenamefont
  {Yoshida}, \citenamefont {Hinohara},\ and\ \citenamefont
  {Nakatsukasa}}]{yos11}%
  \BibitemOpen
  \bibfield  {author} {\bibinfo {author} {\bibfnamefont {K.}~\bibnamefont
  {Yoshida}}, \bibinfo {author} {\bibfnamefont {N.}~\bibnamefont {Hinohara}}, \
  and\ \bibinfo {author} {\bibfnamefont {T.}~\bibnamefont {Nakatsukasa}},\
  }\href {\doibase 10.1088/1742-6596/321/1/012017} {\bibfield  {journal}
  {\bibinfo  {journal} {J. Phys. Conf. Ser.}\ }\textbf {\bibinfo {volume}
  {321}},\ \bibinfo {pages} {012017} (\bibinfo {year} {2011})}\BibitemShut
  {NoStop}%
\bibitem [{\citenamefont {Martini}\ \emph {et~al.}(2016)\citenamefont
  {Martini}, \citenamefont {P\'eru}, \citenamefont {Hilaire}, \citenamefont
  {Goriely},\ and\ \citenamefont {Lechaftois}}]{mar16a}%
  \BibitemOpen
  \bibfield  {author} {\bibinfo {author} {\bibfnamefont {M.}~\bibnamefont
  {Martini}}, \bibinfo {author} {\bibfnamefont {S.}~\bibnamefont {P\'eru}},
  \bibinfo {author} {\bibfnamefont {S.}~\bibnamefont {Hilaire}}, \bibinfo
  {author} {\bibfnamefont {S.}~\bibnamefont {Goriely}}, \ and\ \bibinfo
  {author} {\bibfnamefont {F.}~\bibnamefont {Lechaftois}},\ }\href {\doibase
  10.1103/PhysRevC.94.014304} {\bibfield  {journal} {\bibinfo  {journal} {Phys.
  Rev. C}\ }\textbf {\bibinfo {volume} {94}},\ \bibinfo {pages} {014304}
  (\bibinfo {year} {2016})},\ \Eprint {http://arxiv.org/abs/1607.08483}
  {arXiv:1607.08483 [nucl-th]} \BibitemShut {NoStop}%
\bibitem [{\citenamefont {Doornenbal}\ \emph {et~al.}(2013)\citenamefont
  {Doornenbal}, \citenamefont {Scheit}, \citenamefont {Takeuchi}, \citenamefont
  {Aoi}, \citenamefont {Li}, \citenamefont {Matsushita}, \citenamefont
  {Steppenbeck}, \citenamefont {Wang}, \citenamefont {Baba}, \citenamefont
  {Crawford}, \citenamefont {Hoffman}, \citenamefont {Hughes}, \citenamefont
  {Ideguchi}, \citenamefont {Kobayashi}, \citenamefont {Kondo}, \citenamefont
  {Lee}, \citenamefont {Michimasa}, \citenamefont {Motobayashi}, \citenamefont
  {Sakurai}, \citenamefont {Takechi}, \citenamefont {Togano}, \citenamefont
  {Winkler},\ and\ \citenamefont {Yoneda}}]{doo13}%
  \BibitemOpen
  \bibfield  {author} {\bibinfo {author} {\bibfnamefont {P.}~\bibnamefont
  {Doornenbal}}, \bibinfo {author} {\bibfnamefont {H.}~\bibnamefont {Scheit}},
  \bibinfo {author} {\bibfnamefont {S.}~\bibnamefont {Takeuchi}}, \bibinfo
  {author} {\bibfnamefont {N.}~\bibnamefont {Aoi}}, \bibinfo {author}
  {\bibfnamefont {K.}~\bibnamefont {Li}}, \bibinfo {author} {\bibfnamefont
  {M.}~\bibnamefont {Matsushita}}, \bibinfo {author} {\bibfnamefont
  {D.}~\bibnamefont {Steppenbeck}}, \bibinfo {author} {\bibfnamefont
  {H.}~\bibnamefont {Wang}}, \bibinfo {author} {\bibfnamefont {H.}~\bibnamefont
  {Baba}}, \bibinfo {author} {\bibfnamefont {H.}~\bibnamefont {Crawford}},
  \bibinfo {author} {\bibfnamefont {C.~R.}\ \bibnamefont {Hoffman}}, \bibinfo
  {author} {\bibfnamefont {R.}~\bibnamefont {Hughes}}, \bibinfo {author}
  {\bibfnamefont {E.}~\bibnamefont {Ideguchi}}, \bibinfo {author}
  {\bibfnamefont {N.}~\bibnamefont {Kobayashi}}, \bibinfo {author}
  {\bibfnamefont {Y.}~\bibnamefont {Kondo}}, \bibinfo {author} {\bibfnamefont
  {J.}~\bibnamefont {Lee}}, \bibinfo {author} {\bibfnamefont {S.}~\bibnamefont
  {Michimasa}}, \bibinfo {author} {\bibfnamefont {T.}~\bibnamefont
  {Motobayashi}}, \bibinfo {author} {\bibfnamefont {H.}~\bibnamefont
  {Sakurai}}, \bibinfo {author} {\bibfnamefont {M.}~\bibnamefont {Takechi}},
  \bibinfo {author} {\bibfnamefont {Y.}~\bibnamefont {Togano}}, \bibinfo
  {author} {\bibfnamefont {R.}~\bibnamefont {Winkler}}, \ and\ \bibinfo
  {author} {\bibfnamefont {K.}~\bibnamefont {Yoneda}},\ }\href {\doibase
  10.1103/PhysRevLett.111.212502} {\bibfield  {journal} {\bibinfo  {journal}
  {Phys. Rev. Lett.}\ }\textbf {\bibinfo {volume} {111}},\ \bibinfo {pages}
  {212502} (\bibinfo {year} {2013})}\BibitemShut {NoStop}%
\bibitem [{\citenamefont {Crawford}\ \emph {et~al.}(2019)\citenamefont
  {Crawford}, \citenamefont {Fallon}, \citenamefont {Macchiavelli},
  \citenamefont {Doornenbal}, \citenamefont {Aoi}, \citenamefont {Browne},
  \citenamefont {Campbell}, \citenamefont {Chen}, \citenamefont {Clark},
  \citenamefont {Cort\'es}, \citenamefont {Cromaz}, \citenamefont {Ideguchi},
  \citenamefont {Jones}, \citenamefont {Kanungo}, \citenamefont {MacCormick},
  \citenamefont {Momiyama}, \citenamefont {Murray}, \citenamefont {Niikura},
  \citenamefont {Paschalis}, \citenamefont {Petri}, \citenamefont {Sakurai},
  \citenamefont {Salathe}, \citenamefont {Schrock}, \citenamefont
  {Steppenbeck}, \citenamefont {Takeuchi}, \citenamefont {Tanaka},
  \citenamefont {Taniuchi}, \citenamefont {Wang},\ and\ \citenamefont
  {Wimmer}}]{cra19}%
  \BibitemOpen
  \bibfield  {author} {\bibinfo {author} {\bibfnamefont {H.~L.}\ \bibnamefont
  {Crawford}}, \bibinfo {author} {\bibfnamefont {P.}~\bibnamefont {Fallon}},
  \bibinfo {author} {\bibfnamefont {A.~O.}\ \bibnamefont {Macchiavelli}},
  \bibinfo {author} {\bibfnamefont {P.}~\bibnamefont {Doornenbal}}, \bibinfo
  {author} {\bibfnamefont {N.}~\bibnamefont {Aoi}}, \bibinfo {author}
  {\bibfnamefont {F.}~\bibnamefont {Browne}}, \bibinfo {author} {\bibfnamefont
  {C.~M.}\ \bibnamefont {Campbell}}, \bibinfo {author} {\bibfnamefont
  {S.}~\bibnamefont {Chen}}, \bibinfo {author} {\bibfnamefont {R.~M.}\
  \bibnamefont {Clark}}, \bibinfo {author} {\bibfnamefont {M.~L.}\ \bibnamefont
  {Cort\'es}}, \bibinfo {author} {\bibfnamefont {M.}~\bibnamefont {Cromaz}},
  \bibinfo {author} {\bibfnamefont {E.}~\bibnamefont {Ideguchi}}, \bibinfo
  {author} {\bibfnamefont {M.~D.}\ \bibnamefont {Jones}}, \bibinfo {author}
  {\bibfnamefont {R.}~\bibnamefont {Kanungo}}, \bibinfo {author} {\bibfnamefont
  {M.}~\bibnamefont {MacCormick}}, \bibinfo {author} {\bibfnamefont
  {S.}~\bibnamefont {Momiyama}}, \bibinfo {author} {\bibfnamefont
  {I.}~\bibnamefont {Murray}}, \bibinfo {author} {\bibfnamefont
  {M.}~\bibnamefont {Niikura}}, \bibinfo {author} {\bibfnamefont
  {S.}~\bibnamefont {Paschalis}}, \bibinfo {author} {\bibfnamefont
  {M.}~\bibnamefont {Petri}}, \bibinfo {author} {\bibfnamefont
  {H.}~\bibnamefont {Sakurai}}, \bibinfo {author} {\bibfnamefont
  {M.}~\bibnamefont {Salathe}}, \bibinfo {author} {\bibfnamefont
  {P.}~\bibnamefont {Schrock}}, \bibinfo {author} {\bibfnamefont
  {D.}~\bibnamefont {Steppenbeck}}, \bibinfo {author} {\bibfnamefont
  {S.}~\bibnamefont {Takeuchi}}, \bibinfo {author} {\bibfnamefont {Y.~K.}\
  \bibnamefont {Tanaka}}, \bibinfo {author} {\bibfnamefont {R.}~\bibnamefont
  {Taniuchi}}, \bibinfo {author} {\bibfnamefont {H.}~\bibnamefont {Wang}}, \
  and\ \bibinfo {author} {\bibfnamefont {K.}~\bibnamefont {Wimmer}},\ }\href
  {\doibase 10.1103/PhysRevLett.122.052501} {\bibfield  {journal} {\bibinfo
  {journal} {Phys. Rev. Lett.}\ }\textbf {\bibinfo {volume} {122}},\ \bibinfo
  {pages} {052501} (\bibinfo {year} {2019})}\BibitemShut {NoStop}%
\bibitem [{\citenamefont {Terasaki}\ \emph {et~al.}(1997)\citenamefont
  {Terasaki}, \citenamefont {Flocard}, \citenamefont {Heenen},\ and\
  \citenamefont {Bonche}}]{ter97}%
  \BibitemOpen
  \bibfield  {author} {\bibinfo {author} {\bibfnamefont {J.}~\bibnamefont
  {Terasaki}}, \bibinfo {author} {\bibfnamefont {H.}~\bibnamefont {Flocard}},
  \bibinfo {author} {\bibfnamefont {P.-H.}\ \bibnamefont {Heenen}}, \ and\
  \bibinfo {author} {\bibfnamefont {P.}~\bibnamefont {Bonche}},\ }\href
  {\doibase https://doi.org/10.1016/S0375-9474(97)00183-8} {\bibfield
  {journal} {\bibinfo  {journal} {Nuclear Physics A}\ }\textbf {\bibinfo
  {volume} {621}},\ \bibinfo {pages} {706 } (\bibinfo {year} {1997})},\ \Eprint
  {http://arxiv.org/abs/nucl-th/9612058} {arXiv:nucl-th/9612058} \BibitemShut
  {NoStop}%
\bibitem [{\citenamefont {Stoitsov}\ \emph {et~al.}(2003)\citenamefont
  {Stoitsov}, \citenamefont {Dobaczewski}, \citenamefont {Nazarewicz},
  \citenamefont {Pittel},\ and\ \citenamefont {Dean}}]{sto03}%
  \BibitemOpen
  \bibfield  {author} {\bibinfo {author} {\bibfnamefont {M.~V.}\ \bibnamefont
  {Stoitsov}}, \bibinfo {author} {\bibfnamefont {J.}~\bibnamefont
  {Dobaczewski}}, \bibinfo {author} {\bibfnamefont {W.}~\bibnamefont
  {Nazarewicz}}, \bibinfo {author} {\bibfnamefont {S.}~\bibnamefont {Pittel}},
  \ and\ \bibinfo {author} {\bibfnamefont {D.~J.}\ \bibnamefont {Dean}},\
  }\href {\doibase 10.1103/PhysRevC.68.054312} {\bibfield  {journal} {\bibinfo
  {journal} {Phys. Rev. C}\ }\textbf {\bibinfo {volume} {68}},\ \bibinfo
  {pages} {054312} (\bibinfo {year} {2003})}\BibitemShut {NoStop}%
\bibitem [{\citenamefont {Yoshida}(2009{\natexlab{a}})}]{yos09a}%
  \BibitemOpen
  \bibfield  {author} {\bibinfo {author} {\bibfnamefont {K.}~\bibnamefont
  {Yoshida}},\ }\href {\doibase 10.1140/epja/i2008-10742-y} {\bibfield
  {journal} {\bibinfo  {journal} {Eur. Phys. J. A}\ }\textbf {\bibinfo {volume}
  {42}},\ \bibinfo {pages} {583} (\bibinfo {year} {2009}{\natexlab{a}})},\
  \Eprint {http://arxiv.org/abs/0902.3053} {arXiv:0902.3053 [nucl-th]}
  \BibitemShut {NoStop}%
\bibitem [{\citenamefont {Pei}\ \emph {et~al.}(2009)\citenamefont {Pei},
  \citenamefont {Nazarewicz},\ and\ \citenamefont {Stoitsov}}]{pei09}%
  \BibitemOpen
  \bibfield  {author} {\bibinfo {author} {\bibfnamefont {J.}~\bibnamefont
  {Pei}}, \bibinfo {author} {\bibfnamefont {W.}~\bibnamefont {Nazarewicz}}, \
  and\ \bibinfo {author} {\bibfnamefont {M.}~\bibnamefont {Stoitsov}},\ }\href
  {\doibase 10.1140/epja/i2009-10797-2} {\bibfield  {journal} {\bibinfo
  {journal} {Eur. Phys. J. A}\ }\textbf {\bibinfo {volume} {42}},\ \bibinfo
  {pages} {595} (\bibinfo {year} {2009})},\ \Eprint
  {http://arxiv.org/abs/0901.0545} {arXiv:0901.0545 [nucl-th]} \BibitemShut
  {NoStop}%
\bibitem [{\citenamefont {Yamagami}(2019)}]{yam19}%
  \BibitemOpen
  \bibfield  {author} {\bibinfo {author} {\bibfnamefont {M.}~\bibnamefont
  {Yamagami}},\ }\href {\doibase 10.1103/PhysRevC.100.054302} {\bibfield
  {journal} {\bibinfo  {journal} {Phys. Rev. C}\ }\textbf {\bibinfo {volume}
  {100}},\ \bibinfo {pages} {054302} (\bibinfo {year} {2019})}\BibitemShut
  {NoStop}%
\bibitem [{\citenamefont {Rodr\'iguez-Guzm\'an}\ \emph
  {et~al.}(2002)\citenamefont {Rodr\'iguez-Guzm\'an}, \citenamefont {Egido},\
  and\ \citenamefont {Robledo}}]{rod02}%
  \BibitemOpen
  \bibfield  {author} {\bibinfo {author} {\bibfnamefont {R.}~\bibnamefont
  {Rodr\'iguez-Guzm\'an}}, \bibinfo {author} {\bibfnamefont {J.}~\bibnamefont
  {Egido}}, \ and\ \bibinfo {author} {\bibfnamefont {L.}~\bibnamefont
  {Robledo}},\ }\href {\doibase https://doi.org/10.1016/S0375-9474(02)01019-9}
  {\bibfield  {journal} {\bibinfo  {journal} {Nuclear Physics A}\ }\textbf
  {\bibinfo {volume} {709}},\ \bibinfo {pages} {201 } (\bibinfo {year}
  {2002})},\ \Eprint {http://arxiv.org/abs/nucl-th/0204074}
  {arXiv:nucl-th/0204074} \BibitemShut {NoStop}%
\bibitem [{\citenamefont {Li}\ \emph {et~al.}(2012)\citenamefont {Li},
  \citenamefont {Meng}, \citenamefont {Ring}, \citenamefont {Zhao},\ and\
  \citenamefont {Zhou}}]{li12}%
  \BibitemOpen
  \bibfield  {author} {\bibinfo {author} {\bibfnamefont {L.}~\bibnamefont
  {Li}}, \bibinfo {author} {\bibfnamefont {J.}~\bibnamefont {Meng}}, \bibinfo
  {author} {\bibfnamefont {P.}~\bibnamefont {Ring}}, \bibinfo {author}
  {\bibfnamefont {E.-G.}\ \bibnamefont {Zhao}}, \ and\ \bibinfo {author}
  {\bibfnamefont {S.-G.}\ \bibnamefont {Zhou}},\ }\href {\doibase
  10.1103/PhysRevC.85.024312} {\bibfield  {journal} {\bibinfo  {journal} {Phys.
  Rev. C}\ }\textbf {\bibinfo {volume} {85}},\ \bibinfo {pages} {024312}
  (\bibinfo {year} {2012})},\ \Eprint {http://arxiv.org/abs/1202.0070}
  {arXiv:1202.0070 [nucl-th]} \BibitemShut {NoStop}%
\bibitem [{\citenamefont {Yoshida}(2009{\natexlab{b}})}]{yos09}%
  \BibitemOpen
  \bibfield  {author} {\bibinfo {author} {\bibfnamefont {K.}~\bibnamefont
  {Yoshida}},\ }\href {\doibase 10.1103/PhysRevC.80.044324} {\bibfield
  {journal} {\bibinfo  {journal} {Phys. Rev. C}\ }\textbf {\bibinfo {volume}
  {80}},\ \bibinfo {pages} {044324} (\bibinfo {year} {2009}{\natexlab{b}})},\
  \Eprint {http://arxiv.org/abs/0908.3085} {arXiv:0908.3085 [nucl-th]}
  \BibitemShut {NoStop}%
\bibitem [{\citenamefont {Wang}\ \emph {et~al.}(2017)\citenamefont {Wang},
  \citenamefont {Kortelainen},\ and\ \citenamefont {Pei}}]{wan17b}%
  \BibitemOpen
  \bibfield  {author} {\bibinfo {author} {\bibfnamefont {K.}~\bibnamefont
  {Wang}}, \bibinfo {author} {\bibfnamefont {M.}~\bibnamefont {Kortelainen}}, \
  and\ \bibinfo {author} {\bibfnamefont {J.}~\bibnamefont {Pei}},\ }\href
  {\doibase 10.1103/PhysRevC.96.031301} {\bibfield  {journal} {\bibinfo
  {journal} {Phys. Rev. C}\ }\textbf {\bibinfo {volume} {96}},\ \bibinfo
  {pages} {031301} (\bibinfo {year} {2017})},\ \Eprint
  {http://arxiv.org/abs/1612.06019} {arXiv:1612.06019 [nucl-th]} \BibitemShut
  {NoStop}%
\bibitem [{\citenamefont {Bertsch}\ \emph {et~al.}(1981)\citenamefont
  {Bertsch}, \citenamefont {Cha},\ and\ \citenamefont {Toki}}]{ber81}%
  \BibitemOpen
  \bibfield  {author} {\bibinfo {author} {\bibfnamefont {G.}~\bibnamefont
  {Bertsch}}, \bibinfo {author} {\bibfnamefont {D.}~\bibnamefont {Cha}}, \ and\
  \bibinfo {author} {\bibfnamefont {H.}~\bibnamefont {Toki}},\ }\href {\doibase
  10.1103/PhysRevC.24.533} {\bibfield  {journal} {\bibinfo  {journal} {Phys.
  Rev. C}\ }\textbf {\bibinfo {volume} {24}},\ \bibinfo {pages} {533} (\bibinfo
  {year} {1981})}\BibitemShut {NoStop}%
\bibitem [{\citenamefont {Suzuki}\ and\ \citenamefont {Sagawa}(2000)}]{suz00}%
  \BibitemOpen
  \bibfield  {author} {\bibinfo {author} {\bibfnamefont {T.}~\bibnamefont
  {Suzuki}}\ and\ \bibinfo {author} {\bibfnamefont {H.}~\bibnamefont
  {Sagawa}},\ }\href {\doibase 10.1007/s100500070054} {\bibfield  {journal}
  {\bibinfo  {journal} {Eur. Phys. J. A}\ }\textbf {\bibinfo {volume} {9}},\
  \bibinfo {pages} {49} (\bibinfo {year} {2000})}\BibitemShut {NoStop}%
\bibitem [{\citenamefont {Sagawa}\ \emph {et~al.}(2007)\citenamefont {Sagawa},
  \citenamefont {Yoshida}, \citenamefont {Zhou}, \citenamefont {Yako},\ and\
  \citenamefont {Sakai}}]{sag07}%
  \BibitemOpen
  \bibfield  {author} {\bibinfo {author} {\bibfnamefont {H.}~\bibnamefont
  {Sagawa}}, \bibinfo {author} {\bibfnamefont {S.}~\bibnamefont {Yoshida}},
  \bibinfo {author} {\bibfnamefont {X.-R.}\ \bibnamefont {Zhou}}, \bibinfo
  {author} {\bibfnamefont {K.}~\bibnamefont {Yako}}, \ and\ \bibinfo {author}
  {\bibfnamefont {H.}~\bibnamefont {Sakai}},\ }\href {\doibase
  10.1103/PhysRevC.76.024301} {\bibfield  {journal} {\bibinfo  {journal} {Phys.
  Rev. C}\ }\textbf {\bibinfo {volume} {76}},\ \bibinfo {pages} {024301}
  (\bibinfo {year} {2007})},\ \Eprint {http://arxiv.org/abs/0706.2730}
  {arXiv:0706.2730 [nucl-th]} \BibitemShut {NoStop}%
\bibitem [{\citenamefont {Fracasso}\ and\ \citenamefont
  {Col\`o}(2007)}]{fra07}%
  \BibitemOpen
  \bibfield  {author} {\bibinfo {author} {\bibfnamefont {S.}~\bibnamefont
  {Fracasso}}\ and\ \bibinfo {author} {\bibfnamefont {G.}~\bibnamefont
  {Col\`o}},\ }\href {\doibase 10.1103/PhysRevC.76.044307} {\bibfield
  {journal} {\bibinfo  {journal} {Phys. Rev. C}\ }\textbf {\bibinfo {volume}
  {76}},\ \bibinfo {pages} {044307} (\bibinfo {year} {2007})},\ \Eprint
  {http://arxiv.org/abs/0704.2892} {arXiv:0704.2892 [nucl-th]} \BibitemShut
  {NoStop}%
\bibitem [{\citenamefont {Liang}\ \emph {et~al.}(2012)\citenamefont {Liang},
  \citenamefont {Zhao},\ and\ \citenamefont {Meng}}]{lia12}%
  \BibitemOpen
  \bibfield  {author} {\bibinfo {author} {\bibfnamefont {H.}~\bibnamefont
  {Liang}}, \bibinfo {author} {\bibfnamefont {P.}~\bibnamefont {Zhao}}, \ and\
  \bibinfo {author} {\bibfnamefont {J.}~\bibnamefont {Meng}},\ }\href {\doibase
  10.1103/PhysRevC.85.064302} {\bibfield  {journal} {\bibinfo  {journal} {Phys.
  Rev. C}\ }\textbf {\bibinfo {volume} {85}},\ \bibinfo {pages} {064302}
  (\bibinfo {year} {2012})},\ \Eprint {http://arxiv.org/abs/1201.0071}
  {arXiv:1201.0071 [nucl-th]} \BibitemShut {NoStop}%
\bibitem [{\citenamefont {Akimune}\ \emph {et~al.}(1994)\citenamefont
  {Akimune}, \citenamefont {Daito}, \citenamefont {Fujita}, \citenamefont
  {Fujiwara}, \citenamefont {Greenfield}, \citenamefont {Harakeh},
  \citenamefont {Inomata}, \citenamefont {J\"{a}necke}, \citenamefont {Katori},
  \citenamefont {Nakayama}, \citenamefont {Sakai}, \citenamefont {Sakemi},
  \citenamefont {Tanaka},\ and\ \citenamefont {Yosoi}}]{aki94}%
  \BibitemOpen
  \bibfield  {author} {\bibinfo {author} {\bibfnamefont {H.}~\bibnamefont
  {Akimune}}, \bibinfo {author} {\bibfnamefont {I.}~\bibnamefont {Daito}},
  \bibinfo {author} {\bibfnamefont {Y.}~\bibnamefont {Fujita}}, \bibinfo
  {author} {\bibfnamefont {M.}~\bibnamefont {Fujiwara}}, \bibinfo {author}
  {\bibfnamefont {M.}~\bibnamefont {Greenfield}}, \bibinfo {author}
  {\bibfnamefont {M.}~\bibnamefont {Harakeh}}, \bibinfo {author} {\bibfnamefont
  {T.}~\bibnamefont {Inomata}}, \bibinfo {author} {\bibfnamefont
  {J.}~\bibnamefont {J\"{a}necke}}, \bibinfo {author} {\bibfnamefont
  {K.}~\bibnamefont {Katori}}, \bibinfo {author} {\bibfnamefont
  {S.}~\bibnamefont {Nakayama}}, \bibinfo {author} {\bibfnamefont
  {H.}~\bibnamefont {Sakai}}, \bibinfo {author} {\bibfnamefont
  {Y.}~\bibnamefont {Sakemi}}, \bibinfo {author} {\bibfnamefont
  {M.}~\bibnamefont {Tanaka}}, \ and\ \bibinfo {author} {\bibfnamefont
  {M.}~\bibnamefont {Yosoi}},\ }\href {\doibase
  https://doi.org/10.1016/0375-9474(94)90115-5} {\bibfield  {journal} {\bibinfo
   {journal} {Nucl. Phys. A}\ }\textbf {\bibinfo {volume} {569}},\ \bibinfo
  {pages} {245 } (\bibinfo {year} {1994})}\BibitemShut {NoStop}%
\bibitem [{\citenamefont {Okamura}\ \emph {et~al.}(1995)\citenamefont
  {Okamura}, \citenamefont {Fujita}, \citenamefont {Hara}, \citenamefont
  {Hatanaka}, \citenamefont {Ichihara}, \citenamefont {Ishida}, \citenamefont
  {Katoh}, \citenamefont {Niizeki}, \citenamefont {Ohnuma}, \citenamefont
  {Otsu}, \citenamefont {Sakai}, \citenamefont {Sakamoto}, \citenamefont
  {Satou}, \citenamefont {Uesaka}, \citenamefont {Wakasa},\ and\ \citenamefont
  {Yamashita}}]{oka95}%
  \BibitemOpen
  \bibfield  {author} {\bibinfo {author} {\bibfnamefont {H.}~\bibnamefont
  {Okamura}}, \bibinfo {author} {\bibfnamefont {S.}~\bibnamefont {Fujita}},
  \bibinfo {author} {\bibfnamefont {Y.}~\bibnamefont {Hara}}, \bibinfo {author}
  {\bibfnamefont {K.}~\bibnamefont {Hatanaka}}, \bibinfo {author}
  {\bibfnamefont {T.}~\bibnamefont {Ichihara}}, \bibinfo {author}
  {\bibfnamefont {S.}~\bibnamefont {Ishida}}, \bibinfo {author} {\bibfnamefont
  {K.}~\bibnamefont {Katoh}}, \bibinfo {author} {\bibfnamefont
  {T.}~\bibnamefont {Niizeki}}, \bibinfo {author} {\bibfnamefont
  {H.}~\bibnamefont {Ohnuma}}, \bibinfo {author} {\bibfnamefont
  {H.}~\bibnamefont {Otsu}}, \bibinfo {author} {\bibfnamefont {H.}~\bibnamefont
  {Sakai}}, \bibinfo {author} {\bibfnamefont {N.}~\bibnamefont {Sakamoto}},
  \bibinfo {author} {\bibfnamefont {Y.}~\bibnamefont {Satou}}, \bibinfo
  {author} {\bibfnamefont {T.}~\bibnamefont {Uesaka}}, \bibinfo {author}
  {\bibfnamefont {T.}~\bibnamefont {Wakasa}}, \ and\ \bibinfo {author}
  {\bibfnamefont {T.}~\bibnamefont {Yamashita}},\ }\href {\doibase
  https://doi.org/10.1016/0370-2693(94)01607-E} {\bibfield  {journal} {\bibinfo
   {journal} {Phys. Lett. B}\ }\textbf {\bibinfo {volume} {345}},\ \bibinfo
  {pages} {1 } (\bibinfo {year} {1995})}\BibitemShut {NoStop}%
\bibitem [{\citenamefont {{M.A. {de Huu} and A.M. {van den Berg} and N. Blasi
  and R. {De Leo} and M. Hagemann and M.N. Harakeh and J. Heyse and M. Hunyadi
  and S. Micheletti and H. Okamura and H.J. W\"ortche}}(2007)}]{deh07}%
  \BibitemOpen
  \bibfield  {author} {\bibinfo {author} {\bibnamefont {{M.A. {de Huu} and A.M.
  {van den Berg} and N. Blasi and R. {De Leo} and M. Hagemann and M.N. Harakeh
  and J. Heyse and M. Hunyadi and S. Micheletti and H. Okamura and H.J.
  W\"ortche}}},\ }\href {\doibase
  https://doi.org/10.1016/j.physletb.2007.03.031} {\bibfield  {journal}
  {\bibinfo  {journal} {Phys. Lett. B}\ }\textbf {\bibinfo {volume} {649}},\
  \bibinfo {pages} {35 } (\bibinfo {year} {2007})}\BibitemShut {NoStop}%
\bibitem [{\citenamefont {Dozono}\ \emph {et~al.}(2008)\citenamefont {Dozono},
  \citenamefont {Wakasa}, \citenamefont {Ihara}, \citenamefont {Asaji},
  \citenamefont {Fujita}, \citenamefont {Hatanaka}, \citenamefont {Ishida},
  \citenamefont {Kaneda}, \citenamefont {Matsubara}, \citenamefont {Nagasue},
  \citenamefont {Noro}, \citenamefont {Sakemi}, \citenamefont {Shimizu},
  \citenamefont {Takeda}, \citenamefont {Tameshige}, \citenamefont {Tamii},\
  and\ \citenamefont {Yamada}}]{doz08}%
  \BibitemOpen
  \bibfield  {author} {\bibinfo {author} {\bibfnamefont {M.}~\bibnamefont
  {Dozono}}, \bibinfo {author} {\bibfnamefont {T.}~\bibnamefont {Wakasa}},
  \bibinfo {author} {\bibfnamefont {E.}~\bibnamefont {Ihara}}, \bibinfo
  {author} {\bibfnamefont {S.}~\bibnamefont {Asaji}}, \bibinfo {author}
  {\bibfnamefont {K.}~\bibnamefont {Fujita}}, \bibinfo {author} {\bibfnamefont
  {K.}~\bibnamefont {Hatanaka}}, \bibinfo {author} {\bibfnamefont
  {T.}~\bibnamefont {Ishida}}, \bibinfo {author} {\bibfnamefont
  {T.}~\bibnamefont {Kaneda}}, \bibinfo {author} {\bibfnamefont
  {H.}~\bibnamefont {Matsubara}}, \bibinfo {author} {\bibfnamefont
  {Y.}~\bibnamefont {Nagasue}}, \bibinfo {author} {\bibfnamefont
  {T.}~\bibnamefont {Noro}}, \bibinfo {author} {\bibfnamefont {Y.}~\bibnamefont
  {Sakemi}}, \bibinfo {author} {\bibfnamefont {Y.}~\bibnamefont {Shimizu}},
  \bibinfo {author} {\bibfnamefont {H.}~\bibnamefont {Takeda}}, \bibinfo
  {author} {\bibfnamefont {Y.}~\bibnamefont {Tameshige}}, \bibinfo {author}
  {\bibfnamefont {A.}~\bibnamefont {Tamii}}, \ and\ \bibinfo {author}
  {\bibfnamefont {Y.}~\bibnamefont {Yamada}},\ }\href {\doibase
  10.1143/JPSJ.77.014201} {\bibfield  {journal} {\bibinfo  {journal} {J. Phys.
  Soc. Jap.}\ }\textbf {\bibinfo {volume} {77}},\ \bibinfo {pages} {014201}
  (\bibinfo {year} {2008})}\BibitemShut {NoStop}%
\bibitem [{\citenamefont {Dozono}\ \emph {et~al.}(2020)\citenamefont {Dozono},
  \citenamefont {Uesaka}, \citenamefont {Fukuda}, \citenamefont {Ichimura},
  \citenamefont {Inabe}, \citenamefont {Kawase}, \citenamefont {Kisamori},
  \citenamefont {Kiyokawa}, \citenamefont {Kobayashi}, \citenamefont
  {Kobayashi}, \citenamefont {Kubo}, \citenamefont {Kubota}, \citenamefont
  {Lee}, \citenamefont {Matsushita}, \citenamefont {Michimasa}, \citenamefont
  {Miya}, \citenamefont {Ohkura}, \citenamefont {Ota}, \citenamefont {Sagawa},
  \citenamefont {Sakaguchi}, \citenamefont {Sakai}, \citenamefont {Sasano},
  \citenamefont {Shimoura}, \citenamefont {Shindo}, \citenamefont {Stuhl},
  \citenamefont {Suzuki}, \citenamefont {Tabata}, \citenamefont {Takaki},
  \citenamefont {Takeda}, \citenamefont {Tokieda}, \citenamefont {Wakasa},
  \citenamefont {Yako}, \citenamefont {Yanagisawa}, \citenamefont {Yasuda},
  \citenamefont {Yokoyama}, \citenamefont {Yoshida},\ and\ \citenamefont
  {Zenihiro}}]{doz20}%
  \BibitemOpen
  \bibfield  {author} {\bibinfo {author} {\bibfnamefont {M.}~\bibnamefont
  {Dozono}}, \bibinfo {author} {\bibfnamefont {T.}~\bibnamefont {Uesaka}},
  \bibinfo {author} {\bibfnamefont {N.}~\bibnamefont {Fukuda}}, \bibinfo
  {author} {\bibfnamefont {M.}~\bibnamefont {Ichimura}}, \bibinfo {author}
  {\bibfnamefont {N.}~\bibnamefont {Inabe}}, \bibinfo {author} {\bibfnamefont
  {S.}~\bibnamefont {Kawase}}, \bibinfo {author} {\bibfnamefont
  {K.}~\bibnamefont {Kisamori}}, \bibinfo {author} {\bibfnamefont
  {Y.}~\bibnamefont {Kiyokawa}}, \bibinfo {author} {\bibfnamefont
  {K.}~\bibnamefont {Kobayashi}}, \bibinfo {author} {\bibfnamefont
  {M.}~\bibnamefont {Kobayashi}}, \bibinfo {author} {\bibfnamefont
  {T.}~\bibnamefont {Kubo}}, \bibinfo {author} {\bibfnamefont {Y.}~\bibnamefont
  {Kubota}}, \bibinfo {author} {\bibfnamefont {C.~S.}\ \bibnamefont {Lee}},
  \bibinfo {author} {\bibfnamefont {M.}~\bibnamefont {Matsushita}}, \bibinfo
  {author} {\bibfnamefont {S.}~\bibnamefont {Michimasa}}, \bibinfo {author}
  {\bibfnamefont {H.}~\bibnamefont {Miya}}, \bibinfo {author} {\bibfnamefont
  {A.}~\bibnamefont {Ohkura}}, \bibinfo {author} {\bibfnamefont
  {S.}~\bibnamefont {Ota}}, \bibinfo {author} {\bibfnamefont {H.}~\bibnamefont
  {Sagawa}}, \bibinfo {author} {\bibfnamefont {S.}~\bibnamefont {Sakaguchi}},
  \bibinfo {author} {\bibfnamefont {H.}~\bibnamefont {Sakai}}, \bibinfo
  {author} {\bibfnamefont {M.}~\bibnamefont {Sasano}}, \bibinfo {author}
  {\bibfnamefont {S.}~\bibnamefont {Shimoura}}, \bibinfo {author}
  {\bibfnamefont {Y.}~\bibnamefont {Shindo}}, \bibinfo {author} {\bibfnamefont
  {L.}~\bibnamefont {Stuhl}}, \bibinfo {author} {\bibfnamefont
  {H.}~\bibnamefont {Suzuki}}, \bibinfo {author} {\bibfnamefont
  {H.}~\bibnamefont {Tabata}}, \bibinfo {author} {\bibfnamefont
  {M.}~\bibnamefont {Takaki}}, \bibinfo {author} {\bibfnamefont
  {H.}~\bibnamefont {Takeda}}, \bibinfo {author} {\bibfnamefont
  {H.}~\bibnamefont {Tokieda}}, \bibinfo {author} {\bibfnamefont
  {T.}~\bibnamefont {Wakasa}}, \bibinfo {author} {\bibfnamefont
  {K.}~\bibnamefont {Yako}}, \bibinfo {author} {\bibfnamefont {Y.}~\bibnamefont
  {Yanagisawa}}, \bibinfo {author} {\bibfnamefont {J.}~\bibnamefont {Yasuda}},
  \bibinfo {author} {\bibfnamefont {R.}~\bibnamefont {Yokoyama}}, \bibinfo
  {author} {\bibfnamefont {K.}~\bibnamefont {Yoshida}}, \ and\ \bibinfo
  {author} {\bibfnamefont {J.}~\bibnamefont {Zenihiro}},\ }\href@noop {} {\
  (\bibinfo {year} {2020})},\ \Eprint {http://arxiv.org/abs/2007.15225}
  {arXiv:2007.15225 [nucl-ex]} \BibitemShut {NoStop}%
\end{thebibliography}%

\end{document}